\documentstyle[nato]{crckapb}
\begin{opening}
\title{COSMOLOGICAL PERTURBATIONS OF QUANTUM\protect\\
MECHANICAL ORIGIN AND ANISOTROPY\protect\\
OF THE MICROWAVE BACKGROUND RADIATION}
\author{L. P. Grishchuk}
\institute{McDonnell Center for the Space Sciences, Physics Department\\
Washington University, St. Louis, Missouri 63130, U.S.A.\\
and\\
Sternberg Astronomical Institute, Moscow University\\
119899 Moscow, V234, Russia}
\end{opening}

\begin{document}

A theory of quantum-mechanical generation
of cosmological perturbations is
considered. The conclusion of this study is that
if the large-angular-scale anisotropy in the cosmic microwave
background radiation is caused by the long-wavelength
cosmological perturbations of quantum mechanical origin, they
are, most likely, gravitational waves, rather than density
perturbations or rotational perturbations. Some disagreements with
previous publications are clarified.

This contribution to the Proceedings is based on Reference~[34].

\section{Introduction}

The ongoing and planned high-precision measurements of the
anisotropy in the cosmic microwave background radiation (CMBR)~[1]
may have serious impact on our views about the very early Universe.
The large-angular-scale anisotropy in the CMBR is most likely caused
by cosmological perturbations with wavelengths of the order and longer
than the present day Hubble radius $l_H$.  It is reasonable to expect
that so long-wavelength perturbations are ``primordial'', survived
from the epochs when the Universe was much younger. The wavelengths
of the perturbations have enormously grown up since the time
of generation but other physical characteristics of the perturbations
can still carry imprints of their origin. This cannot be said with
the same degree of certainty about the relatively short-wavelength
cosmological perturbations (unless they are gravitational waves)
which could have been distorted and contaminated in course of their
life by many physical processes occurring in the Universe.

It is remarkable that the origin of all three possible types
of cosmological perturbations, that is the origin of density
perturbations, rotational perturbations, and gravitational waves,
may be of purely quantum-mechanical nature. Cosmological perturbations
can be treated as excitations in gravitational field. In case
of gravitational waves, they are just excitations in gravitational
field itself. In case of density perturbations and rotational
perturbations, they are excitations in gravitational field which
accompany excitations in matter. In the very distant past,
the density and rotational perturbations were excitations in
the primeval medium that was filling the Universe at that time.
The quantum-mechanical generation mechanism of cosmological
perturbations relies only upon the existence of their zero-point
quantum fluctuations and the nonvanishing parametric coupling of
the perturbations to the variable gravitational field of
the homogeneous isotropic Universe. The strong variable gravitational
field of the very early Universe played the role of the pump field.
It supplied energy to the zero-point quantum fluctuations and
amplified them. More precisely, the initial vacuum quantum state
of each mode of the perturbations has been transformed,
as a result of the quantum mechanical Schr\"odinger evolution,
into a multiparticle state known as a squeezed vacuum quantum state.
The generated perturbations have formed a collection of standing waves.
The gravitational field of each of the three types of quantum
mechanically generated perturbations can affect the propagating
photons of the CMBR and produce anisotropy in CMBR.

It was already
emphasized~[2] that there is a significant qualitative difference
between gravitational waves on one side and density and rotational
perturbations on the other side, with regard to possibility of their
quantum mechanical generation. Gravitational waves oscillate in
the absence of external gravitational fields, and their appropriate
parametric coupling to the pump field follows directly from
the Einstein equations. The parametric excitation vanishes only
if the cosmological scale factor obeys the equation
$a^{\prime\prime}/a=0$, that is when there is no any pump field
at all, $a(\eta )={\rm const}$, or when the coupling $a^\prime$
is time independent. Up to this exception, one can say that
the quantum mechanical generation of gravitational waves (relic
gravitons) is unavoidable~[3]. As for density and rotational
perturbations, they are perturbations in matter being accompanied
by perturbations of gravitational field. The ability to support
oscillations of density and/or rotation and the form of their
coupling to the pump field depend on a particular model of matter
and its energy-momentum tensor. The very possibility of the quantum
mechanical generation of these perturbations is model dependent.
Recalling Einstein's definition of two pillars supporting general
relativity, one can say that the quantum mechanically generated
gravitational waves are associated with the pillar made of marble,
while density and rotational perturbations are associated with
the other one.

Presently, we do not know if the observed large-scale structure
in the Universe did really develop from the primordial density
perturbations of quantum-mechanical origin. It is conceivable
that the primordial perturbations arose as a result of, say,
a sophisticated nonsimultaneous phase transition in the very
early Universe or as a consequence of the hypothetical cosmic
defects. All reasonable hypotheses should be carefully studied.
It is necessary, however, to see clearly the difference in the
number and value of initial assumptions: the quantum-mechanical
generating mechanism requires only the validity of general relativity
and basic principles of quantum field theory.  Moreover,
the analogous process of parametric amplification and generation
of squeezed vacuum states of light has been demonstrated
in quantum optics laboratory experiments. It seems that
we are given a great opportunity to derive the cosmological
perturbations from the very first principles. This is why the
quantum-mechanical generation mechanism will be the subject
of our study here.

As was discussed above, gravitational waves are generated
universally in contrast to density and rotational perturbations
which require appropriate matter.  A particular sort of matter
that has received much attention in the recent cosmological
literature, especially in the literature on inflation~[4],
is one or another version of a scalar field. Scalar fields
is a nice theoretical model that has been used in physics
in many different studies. Whether the global scalar fields
do really exist in nature and, if so, whether they couple
to gravity in the way we want, is presently unknown.
However, we will follow the modern tradition in theoretical
physics which states that everything that is not forbidden
is allowed. One can at least guarantee that a sort of
inflationary expansion is a typical feature (attracting separatrix)
in the space of homogeneous isotropic solutions to the Einstein
equations with certain scalar fields~[5]. Scalar fields cannot
support rotational perturbations but they can support density
perturbations.

Specifically, we will study a scalar field $\varphi (t,x^1,x^2,x^3)$
with the energy-momentum tensor
\begin{equation}
 T_{\mu\nu} = \varphi_{,\mu} \varphi_{,\nu}-g_{\mu\nu}
 \Biggl[ {1\over 2}\, g^{\alpha\beta}
 \varphi_{,\alpha}\varphi_{,\beta}+V(\varphi ) \Biggr]
\end{equation}
where $V(\varphi )$ is an arbitrary scalar field potential,
comma denotes a partial derivative. This model of the primeval
cosmological medium satisfies both conditions for the quantum
mechanical generation of density perturbations be possible.
First, the field can obviously support free oscillations
in Minkowski space-time. Second, the explicit form of
the energy-momentum tensor (1) reflects the appropriate
(minimal, the same as for gravitational waves) coupling
of the scalar field to gravity which was chosen by our will.
So, on general grounds and by analogy with gravitational waves,
one can expect that some amount of density perturbations
might have been generated by strong variable gravitational
field of the early Universe. The problem is to quantify
this expectation and to derive the observational predictions,
as reliable and detailed as possible, including the expected
variations in the CMBR.

Scalar fields and scalar field perturbations is a very popular
subject in the framework of inflationary cosmologies.
So popular, that many believe that the inflationary type
of expansion is conditioned by the existence of scalar fields
and that the very possibility to generate perturbations quantum
mechanically relies on the existence of the De Sitter event horizon.
This is not so. Inflation, if understood as a statement about
the behavior of the time dependent cosmological scale factor,
and not about creating and resolving the particle physics paradoxes,
is a phenomenon more general than one particular realization
of it with the help of a scalar field.  [The attitude toward the
archetype inflationary solution~---~exponential expansion~---~has
changed over the years. Astronomers of the older generation
were embarrassed with the De Sitter solution but tried to apply it
for the explanation of the galaxies' red shifts and statistics
of quasars in the most recent Universe. Cosmologists of our time
take the exponential expansion as something almost proven
but apply it to the very remote stages of evolution, somewhere
near the Planck time.]\quad And the quantum mechanical generation
of perturbations is a phenomenon more general and universal than
such concepts as global scalar fields, event horizons, and inflation.
If it turns out that the inflationary hypothesis contradicts observations,
the quantum mechanical generating mechanism will not die together
with inflation. There is little doubt, for instance, that the search
for relic gravitational waves will continue, with may be larger
emphasis on relatively short waves rather than on long waves~[6].
And a test of the quantum mechanical origin of cosmological
perturbations will be a test of their origin, not a test
of inflation specifically.

The generation of density perturbations in inflationary models governed
by the scalar field (1) was a subject of discussion in many research and
review articles, and books. If one consults the most recent literature,
one can find that the current situation is often summarized in the
following, or similar, words: ``Exponential inflation predicts a
scale-invariant, Gaussian spectrum of scalar fluctuations~...,
and a smaller amount of tensor fluctuations~...~. Other inflationary
models, for instance power-law inflation~..., predict spectra slightly
tilted away from scale invariance.''\quad  (See, for example,~[7]
and references therein.)\quad  The expected amplitudes of density
perturbations are usually quoted in the following, or equivalent,
form (see, for instance,~[8] and references therein):
\[
\Biggl( {\delta\rho \over \rho} \Biggr)_\lambda^{\rm hor}
= {m\kappa^2\over 8\pi^{3/2}}\,
  {H^2(\varphi )\over |H^\prime (\varphi )|}\quad ,
\]
where the quantities on the right-hand-side are supposed to be evaluated
``when the scale $\lambda$ crossed the Hubble radius during inflation''.
The denominator of this expression depends on the derivative of the Hubble
parameter and goes to zero in the limit of exponential inflation.
Apparently, this formula says that the predicted amplitudes of the
scale-invariant spectrum are arbitrarily close to infinity,
and the amplitudes of nearby spectra are ``slightly tilted away''
from infinity. According to this formula, the amplitudes of density
perturbations are many orders of magnitude larger than the amplitudes
of gravitational waves, if the expansion rate is sufficiently close
to the exponential inflation. The belief that the amplitudes
of density perturbations are larger, or much larger,
than the amplitudes of gravitational waves is considered
to be a strong prediction of inflationary models based
on the scalar field (1). For instance, the author of Ref.~[8]
concludes:\quad ``An observation violating this condition
at any scale would immediately rule out the general class
of models we are considering''. The expected contribution
of density perturbations and gravitational waves
to the quadrupole anisotropy of CMBR was also under study.
The authors of Ref.~[9] (see also [10] and references therein)
say that ``The ratio
of gravitational wave ($T$) to energy-density perturbations ($S$)
contributions to the CMB quadrupole anisotropy is predicted to be
$T/S = 21(1+\gamma )$'', where $\gamma$, in that paper,
is the parameter in the equation of state for matter governing
the inflationary expansion, $p = \gamma\rho$.
According to this formula, $T/S$ vanishes in the limit of
$\gamma =-1$, that is in the limit of strictly exponential (De Sitter)
inflation.  Apparently, this formula for $T/S$ is based on the
authors' assumption that the effectiveness of generation
of density perturbations is the higher the closer the expansion
law to the exponential inflation, and goes to infinity in the limit
of $\gamma =-1$. Apparently, this is why $T/S$ goes to zero in
this limit. The statements about density perturbations are sometimes
characterized as such that have been ``widely studied and there
is broad agreement regarding both methods and results~...''~[11].

I suspect that the present work will not belong to that category
of studies that enjoyed the ``broad agreement''; my conclusions
are considerably different from what was described in the preceeding
paragraph. I will be arguing that there is no linear density perturbations
at all at the purely exponential (De Sitter) inflationary stage
for models governed by the scalar field (1). Density perturbations
can only arise as a result of violation of the purely exponential
expansion and transition to the radiation-dominated stage.
Regardless of how close to zero was the derivative of the
Hubble parameter ``when the scale $\lambda$ crossed the Hubble
radius during inflation'', the today's amplitudes of density
perturbations are finite. The amplitudes of gravitational waves
are typically a little larger than the amplitudes of density
perturbations, at least in the long wavelength limit where
spectra are smooth and have the power-law behavior.
Correspondingly, the contribution of density perturbations to
the quadrupole anisotropy is never much larger than the contribution
of gravitational waves. In fact, it is somewhat smaller in the limit
of long waves.

As for the statistical properties of cosmological perturbations and,
hence, the statistical properties of the CMBR fluctuations caused
by them, it was already emphasized~[12] that they are determined
by the statistics of quantum states being generated, namely by
the statistics of squeezed vacuum quantum states.

In this paper, we will present detailed derivations of equations
and their solutions in order to make it possible for
the interested reader to compare the present calculations with
those of other authors. The structure of the paper is the following.
In Sec.~II we present the general equations for cosmological perturbations.
The equations are applicable for matter with the energy-momentum tensor
of arbitrary form. They use only the defining property of cosmological
perturbations, namely that the perturbations are based on the scalar,
vector, and tensor functions of spatial coordinates. In Sec.~III we apply
these equations specifically to the initial stage
($i$ stage) of cosmological evolution which is assumed to be governed
by the scalar field with the energy-momentum tensor (1). The main purpose is
to study the density perturbations. No assumptions
about a particular form of the scalar field potential $V(\varphi )$
or a particular (for example, inflationary) type of expansion are
being made {\it a priori}. The time-dependent coefficients of
the differential equations for the perturbations are expressed
in terms of the scale factor (and its derivatives) only.
This reflects the underlying interaction of the perturbations
with the variable gravitational pump field. The determination
of all unknown functions describing density perturbations is reduced
to solving a single differential equation which is very similar
to the equation for gravitational waves. The behavior of solutions
during a more or less gradual transition from the $i$ stage to
the radiation-dominated stage ($e$ stage) is studied in Sec.~IV.
In order to deal with simple exact solutions at both stages
we will be interested in a sharp transition from the $i$ stage
to the $e$ stage. In Sec.~V we apply the perturbation equations
to the perfect fluid matter with arbitrary velocity of sound.
We present solutions to these equations at the $e$ stage and
the matter-dominated stage ($m$ stage) in the form convenient
for matching the solutions at all three stages ($i$, $e$, $m$).
In Sec.~VI we join the solutions, find the coefficients which
were undetermined so far, and express the solution at the $m$
stage entirely in terms of the functions (and their first time
derivatives) describing the perturbations at the time of joining
the $i$ and $e$ stages. As a preparation for quantization of
the perturbations, we briefly discuss the density and rotational
perturbations of matter placed in the Minkowski space-time,
that is neglecting gravitational fields (Sec.~VII).  The quantization
of density perturbations is performed in Sec.~VIII. This procedure
essentially repeats the steps which have been previously done
for gravitational waves and rotational perturbations~[12,2].
The quantum mechanically generated perturbations are placed in
squeezed vacuum quantum states. Classically, one can think of
the perturbations as of a stochastic collection of standing waves.
The justification and necessity of the so-called Sakharov's
oscillations in the spectra of density perturbations is discussed.
In order to get the analytic results as detailed as possible,
we specialize the scale factor of the $i$ stage to the $\eta$-time
power laws which include inflationary models. This allows us
to derive concrete power-law spectra of density perturbations
at the $m$-stage. In Sec.~IX we derive an exact formula for
the angular correlation function of the CMBR temperature
variations $\delta T/T$ caused by squeezed density perturbations.
The multipole decomposition of the correlation function begins
from the monopole term. The contributions to the monopole and
dipole terms produced by individual waves with wavelengths
exceeding $l_H$ are strongly suppressed, which is in agreement
with previous results~[13]. Nevertheless, one should be aware
that not only the entire quadrupole but also some little portions
of the measured mean temperature of CMBR and its dipole variation
may be caused by perturbations of quantum mechanical origin.
For one and the same cosmological model, the contribution of
density perturbations to the quadrupole anisotropy of CMBR is
never much larger than the contribution of gravitational waves,
at least for models considered here.

\section{General Equations for Cosmological Perturbations}

The unperturbed spatially-flat FLRW (Friedmann-Lemaitre-Robertson-\protect\\
Walker) cosmological models are described by the metric
\begin{equation}
   ds^2 = -c^2dt^2 + a^2(t)
   \Bigl( {dx^1}^2 + {dx^2}^2 + {dx^3}^2 \Bigr)
 = -a^2(\eta )
   \Bigl( d\eta^2 - {dx^1}^2 - {dx^2}^2 - {dx^3}^2 \Bigr) \, .
\end{equation}
Here we restrict ourselves to study of the spatially-flat models.
This is done for purposes of simplicity only, not because
of any inflationary prejudices. If it turns out that the
density parameter $\Omega$ is, say, $\Omega =0.8$, instead
of $\Omega =1$, the inflationary hypothesis will prove
to be wrong but the quantum-mechanical generating mechanism
will still be perfectly well effective, although with certain
modifications as compared to the spatially-flat case.

The scale factor $a(\eta )$ is governed by matter with the unperturbed
values of energy density $\epsilon_0$, $T_o^o =-\epsilon_0$,
and pressure $p_0$, $T_i^k = p_0\delta_i^k$:
\begin{eqnarray}
   {3\over a^2} \Biggl( {a^\prime \over a}\Biggr) ^2
&  = \kappa\epsilon_0 \nonumber\\
    - {1\over a^2} \Biggl[ 2 \Biggl( {a^\prime \over a}\Biggr) ^\prime
    + \Biggl( {a^\prime \over a} \Biggr) ^2 \Biggl]
&  = \kappa p_0
\end{eqnarray}
where $\kappa =8\pi G/c^4$ and a prime is $d/d\eta$,
$d/d\eta =(a/c)d/dt$. The Hubble parameter is
$H= \dot a /a =ca^\prime /a^2$.

It is convenient to introduce two new functions of the scale
factor:
\begin{equation}
 \alpha (\eta ) = {a^\prime \over a} \, ,\qquad
 \gamma (\eta ) = 1 - {\alpha^\prime \over \alpha^2} \quad .
\end{equation}
In terms of $t$-time the function $\gamma$ is
$\gamma (t)=-(\dot H /H^2)$.  Due to Eqs.~(3) one has
\begin{equation}
  \kappa (\epsilon_0 + p_0) = {2\alpha^2 \over a^2}\, \gamma \quad .
\end{equation}

The function $\gamma (\eta )$ becomes a constant if $a(\eta )$
is governed by matter with the effective equation of state
$p_0 = q\epsilon_0$, where $q={\rm const}$.
The scale factor $a(\eta )$ takes on the $\eta$-time power law behavior
\begin{equation}
   a = l_o | \eta | ^{1\!+\!\beta}
\end{equation}
($\eta$ must be negative for expanding models with
$1\!+\!\beta < 0$) where $l_o$ and $\beta$ are constants.
(This parameterization has been used before [3].)
The constant $l_o$ has the dimensionality of length.
It follows from Eqs.~(3) and (4) that
$\gamma =(2\!+\!\beta )/(1\!+\!\beta )$,
$q(\beta ) =(1\!-\!\beta )/3(1\!+\!\beta )$,
where the parameter $\beta$ can vary in the interval
$-\infty < \beta < \infty$.
In particular, $\gamma = 2$ at the radiation-dominated stage,
and $\gamma = 3/2$ at the matter-dominated stage.
Note that $\gamma =0$ in case of purely exponential (De Sitter)
expansion for which $a(t) \sim e^{Ht}$, $H={\rm const}$,
$a(\eta )=l_o |\eta |^{-1}$, $\beta =-2$.

One can also derive from Eqs.~(3) the relationship
\begin{equation}
    {p_0^\prime \over \epsilon_0^\prime}
  = -1 + {2\over 3} \, \gamma - {\gamma^\prime \over 3\alpha\gamma}
  = -{1\over 3\alpha} (\ln\, a\alpha^2\gamma )^\prime \, .
\end{equation}
The ratio $p_0^\prime /\epsilon_0^\prime$
becomes a constant for the scale factors (6), namely:\protect\\
$p_0^\prime /\epsilon_0^\prime = q(\beta )$.
In particular, $p_0^\prime /\epsilon_0^\prime$
goes to $-$1 in the limit of $\beta =-$2.

Let us now turn to the perturbations of the spatially-flat FLRW models.
Without resticting in any way the physical content of the problem,
it is convenient to work in the class of synchronous coordinate systems.
(At this point the reader may have to be ready to exhibit certain
resistance to the pressure from the proponents of the ``gauge-invariant''
formalisms.) Using the $\eta$-time coordinate, one can write
the general expression for the metric tensor including perturbations as:
\[
 ds^2=-a^2\left[ d\eta^2-(\delta_{ij}+h_{ij})dx^i dx^j \right] \, ,
\]
where the six functions  $h_{ij}$  describe all the independent
components of the perturbed gravitational field.

The construction of cosmological perturbations [14,15] (see also [16])
is based on the classification of the eigenfunctions to the
3-dimensional Laplace operator. Density perturbations
are associated with the scalar functions $Q(x^1,\, x^2,\, x^3)$
satisfying the equation
\begin{equation}
     {Q^{,i}}_{,i} + n^2Q = 0
\end{equation}
valid in three-space
$dl^2 = d{x^1}^2 + d{x^2}^2 + d{x^3}^2$.
For each wave vector ${\bf n}$,
one can choose two linearly independent solutions to Eq.~(8)
in the form $e^{i{\bf nx}}$ and $e^{-i{\bf nx}}$.
{}From a given scalar field $Q$ one can construct a vector field
$Q_{,i}$ and two tensor fields:\quad $\delta_{ik}Q$ and
$Q_{,i,k}=-n_in_kQ$.  For each ${\bf n}$,
the general perturbation of the energy-momentum tensor
and the accompanying perturbation of the gravitational field
can be written as a sum of products of time-dependent amplitudes
and spatial functions introduced above.

The components of the perturbed gravitational field can be written as
\begin{equation}
  h_{ij} = hQ\delta_{ij} + h_l n^{-2} Q_{,i,j}
\end{equation}
where the function $h(\eta )$ represents the scalar (proportional to
$\delta_{ij}Q$) perturbation of the gravitational field while
the function $h_l(\eta )$ represents the longitudinal-longitudinal
(proportional to $n_in_jQ$) perturbation.  The general expression for
${T_\mu}^\nu$ including perturbations can be written as
\begin{eqnarray}
      T_o^o
& = & - \epsilon_0 - {1\over a^2} \, \epsilon_1Q  \, , \quad
        T_i^o = {1\over a^2} \, \xi^\prime Q_{,i} \, , \quad
        T_o^i = -{1\over a^2} \, \xi^\prime Q^{,i} \, , \nonumber\\
       T_i^k
& = & p_0\delta_i^k + {1\over a^2} (p_1+p_l)Q\delta_i^k
      + {1\over a^2} n^{-2}p_l{Q_{,i}}^{,k} \, .
\end{eqnarray}
The form of Eqs.~(9) and (10) is based solely on the definition
of density perturbations and our choice of synchronous coordinate
systems. In all other respects, the representation (9) and (10)
is general. The particular notations for arbitrary functions
describing the perturbations are chosen for later convenience.

The arbitrary functions
$h(\eta )$, $h_l(\eta )$, $\epsilon_1(\eta )$, $p_1(\eta )$
$p_l(\eta )$, $\xi^\prime (\eta )$ should satisfy all together
the perturbed Einstein equations:
\begin{eqnarray}
    3\alpha h^\prime + n^2h - \alpha h_l^\prime
& = \kappa \epsilon_1\\
    h^\prime
& = \kappa \xi^\prime\\
     -h^{\prime\prime} -2\alpha h^\prime
& = \kappa p_1 \\
    {1\over 2} (h_l^{\prime\prime} + 2\alpha h_l^\prime - n^2h)
& = \kappa p_l \, .
\end{eqnarray}
There are too many unknown functions to be found from
Eqs.~(11)-(14). This requires us to specify a model for matter
and its energy-momentum tensor. We will consider three consecutive
stages of expansion:\quad $i$ stage governed by the scalar field (1),
and the subsequent $e$ and $m$ stages governed by perfect fluid
with the energy-momentum tensor
\begin{equation}
     {T_\mu}^\nu
  = (\epsilon + p) u_\mu u ^\nu + p{\delta_\mu}^\nu \, .
\end{equation}
The equation of state at the $e$ and $m$ stages is
$p={1\over 3} \epsilon$ and $p=0$, respectively.

Now we will turn to the rotational perturbations and gravitational
waves.  The rotational perturbations are based on the vector
functions $Q_i$ satisfying the equations
\[
   {Q_{i,k}}^{,k} + n^2 Q_i = 0\, \qquad  {Q^i}_{,i} = 0 \quad .
\]
The nonvanishing components of the perturbed gravitational
field and the nonvanishing components of the perturbed
energy-momentum tensor can be written in the form
\[
  h_{ij} =2h(\eta )A_{ij}\, ,\quad
  T_i^o =-T_o^i=a^{-2} \omega (\eta )Q_i \, , \quad
  T_i^k =2a^{-2} n^{-1} \chi (\eta )A_i^k
\]
where $A_{ij} =-{1\over 2n} (Q_{i,j}+Q_{j,i})$.
The arbitrary functions
$h(\eta )$, $\omega (\eta )$, $\chi (\eta )$
should satisfy together the perturbed Einstein equations:
\begin{eqnarray}
  && h^{\prime\prime} + 2\alpha h^\prime
     = 2\kappa n^{-1} \chi (\eta )\, ,\nonumber\\
  && -{n\over 2} h^\prime = \kappa\, \omega (\eta ) \, .\nonumber
\end{eqnarray}

And, finally, gravitational waves are based on the tensor functions
$Q_{ij}$ satisfying the equations
\[
  {Q_{ij,k}}^{,k} + n^2 Q_{ij} = 0 \, , \qquad
  {Q^{ij}}_{,j} = 0 \, , \qquad
  Q^{ij} \delta_{ij} = 0 \, .
\]
The components of the perturbed gravitational field should obey
the perturbed Einstein equations
\[
   {h_i}^{j^{\prime\prime}} + 2\alpha h_i^{j^\prime}
 + n^2{h_i}^j = 0 \quad .
\]
Note, that the components ${T_i}^k$ of the perturbed energy-momentum
tensor vanish for simple models of matter, such as perfect fluids,
but, in general, they may be different from zero.
However, we put them equal to zero since we are not interested
in the generation of gravitational waves by anisotropic stresses
of the material medium. Instead, we are interested in the
quantum-mechanical generation of gravitational waves by variable
pump field.  This process is independent of the material anisotropic
stresses and takes place even if there is no classical sources at all.

By introducing
\[
    h_{ij} = {1\over a} \, \mu (\eta )Q_{ij}
\]
one can reduce the basic equation to the form of the equation for the
parametrically excited oscillator with the time-dependent frequency
\[
   \mu^{\prime\prime} + \mu
\left( n^2 - {a^{\prime\prime} \over a} \right) = 0
\]
which reflects the underlying physics of the phenomenon.
This equation has been studied before (for a review, see Ref.~[19]).

At the first glance, the original equations for rotational
and density perturbations do not look like equations for
a parametrically coupled oscillator.  However, they can be
reduced to this equation for appropriate models of matter
whenever the amplification process is at all possible.
In case of rotational perturbations this has been done
in Ref.~[2].  For density perturbations we will do it below
by concentrating on the scalar field matter with the
energy-momentum tensor (1).

\section{Density Perturbations at the Initial Stage of Expansion\protect\\
Governed by a Scalar Field}

At the $i$ stage, the evolution of the scale factor $a(\eta )$
is determined by the unperturbed homogeneous scalar field
$\varphi =\varphi_o(\eta )$. The unperturbed
values $\epsilon_0$, $p_0$ are given by Eq.~(1):
\begin{eqnarray}
      \epsilon_0
& = & {1\over 2a^2} (\varphi_0^\prime )^2 + V(\varphi ) \\
      p_0
& = & {1\over 2a^2} (\varphi_0^\prime )- V(\varphi ) \quad .
\end{eqnarray}
By summing up Eqs.~(16) and (17) and comparing the result with Eq.~(5)
one can derive the equation
\begin{equation}
   \kappa (\varphi_0^\prime )^2 = 2\alpha^2\gamma \quad .
\end{equation}
It follows from this equation that  $\gamma \geq 0$
for the scale factors governed by the scalar field (1).
The De Sitter case corresponds to $\varphi_o^\prime =0$,
$\varphi_0={\rm const}$ and
$\epsilon_0 = -p_0 = V(\varphi_0)={\rm const}$.

If $\varphi_0^\prime \neq 0$, one can use the equation
\[
\epsilon_0^\prime =-\!3\alpha(\epsilon_0 +p_0 )
\]
which is a consequence of Eq.~(3), and obtain with the help
of Eqs.~(16) and (17):
\begin{equation}
  \varphi_0^{\prime\prime} +2\alpha\varphi_0^\prime +a^2V_{,\varphi}=0
\end{equation}
where $V_{,\varphi} = dV(\varphi )/d\varphi$, the derivative
is taken at $\varphi=\varphi_0$.  The further useful relationships
following from Eqs.~(18) and (19) are:
\begin{eqnarray}
     {\varphi_0^{\prime\prime} \over \varphi_0^\prime}
& = &{\alpha^\prime \over \alpha}
     + {1\over 2}\, {\gamma^\prime\over\gamma} \\
     -{a^2\over\varphi_0^\prime} \, V_{,\varphi}
& = & 2\alpha + {\alpha^\prime\over\alpha} + {1\over 2}\,
      {\gamma^\prime\over\gamma} \quad .
\end{eqnarray}

The perturbations of the gravitational field are associated
with the perturbations of the scalar field.  We will write
the perturbations of the scalar field as
\begin{equation}
    \varphi = \varphi_0(\eta ) + \varphi_1(\eta )Q \quad .
\end{equation}
Having at our disposition the energy-momentum tensor (1)
and the definitions (22), (9), (10) we can directly calculate
the functions $\epsilon_1$, $\xi^\prime$, $p_1$, $p_l$:
\begin{eqnarray}
      \epsilon_1
& = & \varphi_0^\prime \varphi_1^\prime
      + a^2\varphi_1 V_{,\varphi} \\
      \xi^\prime
& = & -\varphi_0^\prime\varphi_1 \\
      p_1
& = & \varphi_0^\prime\varphi_1^\prime
      - a^2\varphi_1 V_{,\varphi} \\
      p_l
& = & 0 \quad .
\end{eqnarray}

We will now assume that $\varphi_0^\prime \neq 0$.
The De Sitter case $\varphi_0^\prime =0$ will be considered separately
at the end of this Section. It follows from Eq.~(24) that
$\varphi_1 = -\xi^\prime /\varphi_0^\prime$.
Inserting this value of $\varphi_1$ into Eqs.~(23) and (25)
and using Eqs.~(20) and (21), one can express
$\epsilon_1$, $p_1$, in terms of $h(\eta )$:
\begin{eqnarray}
      \kappa \epsilon_1
& = & -h^{\prime\prime} + h^\prime
      \Biggl( 2\alpha + 2{\alpha^\prime\over\alpha}
       + {\gamma^\prime\over\gamma} \Biggr) \\
      \kappa p_1
& = & -h^{\prime\prime} - 2\alpha h^\prime \quad .
\end{eqnarray}

We should now return to the perturbed Einstein equations (11)-(14)
making use of Eqs.~(27) and (28).  Equation~(11) can be written as an
expression for $h_l^\prime (\eta )$ in terms of $h(\eta )$:
\begin{equation}
   h_l^\prime = {1\over \alpha}
\Biggl[ h^{\prime\prime} + h^\prime
\biggl( \alpha -2{\alpha^\prime \over \alpha} -
       {\gamma^\prime \over \gamma} \biggr) + n^2 h \Biggr] \quad .
\end{equation}
Equation (12) expresses $\xi^\prime$ in terms of $h^\prime$.
Equation~(13) is satisfied identically. Equation~(14) reads as
\begin{equation}
     h_l^{\prime\prime} + 2\alpha h_l^\prime -n^2h = 0 \quad .
\end{equation}
Thus, if one knows the function $h(\eta )$, all other functions
describing the density perturbations, namely
$h_l(\eta )$, $p_1(\eta )$, $\epsilon_1(\eta )$,
and $\xi^\prime (\eta )$,
can be found with the help of Eqs.~(29), (28), (27)
(or, equivalently, (11)) and (12). To derive the equation for
$h(\eta )$ we substitute Eq.~(29) into Eq.~(30) and obtain
\begin{equation}
   h^{\prime\prime\prime} + h^{\prime\prime}
\Biggl( 3\alpha\gamma - {\gamma^\prime \over \gamma} \Biggr) + h^\prime
\Biggl[ n^2 -2\alpha^\prime + 2\gamma\alpha^2
      - {\alpha^\prime \over \alpha} {\gamma^\prime\over\gamma}
- \Biggl( {\gamma^\prime\over\gamma} \Biggr)^\prime \Biggr]
  + n^2 \alpha\gamma h = 0 \, .
\end{equation}

Equation (31) is a third-order differential equation.
There should be no wonder (and no panic) on this occasion.
One of solutions to this equation we know in advance,
this solution is
\begin{equation}
   h = C\, {\alpha\over a}
\end{equation}
where $C$ is an arbitrary constant. We could have expected the
existence of this solution, even before solving the equation
for $h(\eta )$, because the perturbation of this form can
be generated by a coordinate transformation which does not
violate our choice of synchronous coordinate systems and,
hence, does not destroy our initial form (9) of the perturbed
metric. (One can easily check that the function (32) is indeed
a solution to Eq.~(31).)  Concretely, one can perform a
small coordinate transformation
\[
  \bar\eta = \eta - {C\over 2a} Q \, ,\quad
  \bar x^i = x^i - {C\over 2} Q^{, i} \int a^{-1} d\eta \quad .
\]
In terms of new coordinates $\bar\eta$, $\bar x^i$
the transformed components (9) take on the form
\[
  \bar g_{oo} = -\!a^2(\bar\eta )\, ,\qquad
  \bar g_{oi} = 0\, ,\qquad
  \bar g_{ik} = a^2(\bar\eta )
  \biggl[ (1+\bar h Q)\delta_{ik}
            -\bar h_ln^{-2} Q_{,i,k} \biggr] \, ,
\]
where
\begin{equation}
  \bar h = h+ C{\alpha\over a} \, , \quad
  \bar h_l = h_l + Cn^2 \int a^{-1}\, d\eta \quad .
\end{equation}
The same transformation should be applied to the components
of the energy-momentum tensor. Even if the original $h$, $h_l$
are zero, the transformed $\bar h$, $\bar h_l$ are not zero.
After erasing the overbars in the transformed components of
the metric, one returns to Eq.~(9). The freedom of choosing
different freely falling coordinate systems and corresponding
spatial slices $\eta ={\rm const}$ gets represented
in the form of freedom to choose different solutions from
the family of all solutions for the perturbations.
All choices of $C$ are equally well ``physical''.
The integral in Eq.~(33) produces an additional integration
constant which reflects the possibility to make a purely
spatial transformation and to shift $h_l$ by a constant value,
but we will not actually need this coordinate freedom.
It follows from Eq.~(33) that there are two functions
(and many algebraic and differential combinations constructed
from them) that do not contain $C$ at all:
\[
   u = h^\prime + \alpha\gamma h\, ,\qquad
   v = h_l^\prime - {1\over\alpha}n^2 h \quad .
\]

The solution (32) allows one to reduce the third-order differential
equation (31) to the second-order differential equation.
To reach this goal we use the function $u(\eta )$:
\begin{equation}
    u = h^\prime + \alpha\gamma h \quad .
\end{equation}
Obviously, this function vanishes on the solution (32).
By substituting Eq.~(34) into Eq.~(31) we derive the equation
for $u(\eta )$:
\begin{equation}
        u^{\prime\prime} + u^\prime
\Biggl( 2\alpha\gamma - {\gamma^\prime\over\gamma} \Biggr) + u
\Biggl[ n^2 -2\alpha^\prime -\alpha {\gamma^\prime\over\gamma} -
\Biggl( {\gamma^\prime\over\gamma} \Biggr)^\prime \Biggr] = 0 \, .
\end{equation}
Note that the coefficients of this differential equation
depend exclusively on the scale factor and its derivatives.
This fact is a manifestation of the underlying interaction
of the perturbations with the cosmological pump field.
No special assumptions about the shape of the potential
$V(\varphi)$ or such things as ``nonsimultaneous rolling of
the scalar field down the hill'' have been made whatsoever.

Our next move is to transform Eq.~(35) to the form similar
to the equation for gravitational waves. This will allow us
to use certain results derived previously for gravitational
waves and rotational perturbations~[17,2].
In order to get rid of $u^\prime$ we introduce the function
$\mu (\eta )$ according to
\begin{equation}
   u = {\alpha\sqrt{\gamma} \over a} \mu \quad .
\end{equation}
It follows from Eq.~(18) that the function $\gamma (\eta )$ is
nonnegative if the scale factor is governed by the scalar field (1)
which we study here. However, Eq.~(35) is formally applicable
to negative $\gamma$ as well. It may happen (as the author thinks) that
Eq.~(35) has a wider domain of validity and can be used,
for other models of matter, with negative $\gamma$ too.
If this is the case, one is free to modify Eq.~(36) by using
$\sqrt{|\gamma |}$ instead of $\sqrt{\gamma}$. Anyway,
with the help of Eq.~(36) one derives the equation
\[
  \mu^{\prime\prime} + \mu [n^2 -U(\eta )] = 0
\]
where
\begin{eqnarray}
      U(\eta )
& = & \alpha^2 + \alpha^\prime + \alpha {\gamma^\prime\over\gamma}
      + {1\over 4} \Biggl( {\gamma^\prime\over\gamma} \Biggr)^2
      + {1\over 2} \Biggl( {\gamma^\prime\over\gamma} \Biggr)^\prime
      = U_0(\eta ) + U_1(\eta ) \, \nonumber\\
      U_0(\eta )
& = & \alpha^2 + \alpha^\prime = {a^{\prime\prime}\over a} \, , \quad
      U_1(\eta ) = {1\over\gamma^2}
      \Biggl[ \alpha\gamma\gamma^\prime
      - {1\over 4} \gamma^{\prime 2}
      + {1\over 2} \gamma\gamma^{\prime\prime} \Biggr] \, .
\end{eqnarray}
The effective potential $U(\eta )$ can also be written as
$U(\eta ) = (a\sqrt{\gamma})^{\prime\prime}/a\sqrt{\gamma}$
which reduces our basic equation to the form
\begin{equation}
    \mu^{\prime\prime} + \mu
\Biggl[ n^2 -
        {(a\sqrt{\gamma})^{\prime\prime}\over a\sqrt{\gamma}}
\Biggr] = 0 \quad .
\end{equation}
(The function $a\sqrt{\gamma}$ can be related with the function
$z$ discussed in [18], see also the early papers [27].)

Let us recall [3] that in the case of gravitational waves the potential
$U(\eta )$ consists only of the $U_0(\eta )$ term,
so that the basic equation is
\begin{equation}
    \mu^{\prime\prime} + \mu
\Biggl[ n^2 - {a^{\prime\prime} \over a} \Biggr] = 0 \quad .
\end{equation}
The gravitational wave potential $U_0$ depends only on
the first and second time derivatives of the logarithm
of the scale factor:
$(\ln\, a)^\prime$,  $(\ln\, a)^{\prime\prime}$.
The potential $U(\eta )$ for density perturbations
is more complicated and includes also
$(\ln\, H)^\prime$, $(\ln\, H)^{\prime\prime}$, and
$(\ln\, H)^{\prime\prime\prime}$.
We note, however, that the potentials are exactly the same,
and, therefore, the basic equations and solutions for density
perturbations and gravitational waves are exactly the same,
if $\gamma$ is constant, that is for the scale factors (6).
For this class of pump fields, the general solution
to Eq.~(39) can be written in the form (for non-half-integer $\beta$):
\begin{equation}
  \mu (\eta ) = (n\eta )^{1/2}
\biggl[ A_1J_{\beta+{1\over 2}} (n\eta )
      + A_2 J_{-(\beta +{1\over 2})} (n\eta )
\biggr] \, .
\end{equation}

Having two linearly independent solutions to Eq.~(38) one can
construct $h(\eta )$ and, hence, to find the rest of functions
describing density perturbations. It follows from Eqs.~(34) and
(36) that
\begin{equation}
  h^\prime = -\gamma \alpha h + {\alpha\sqrt{\gamma} \over a} \mu
\end{equation}
and
\begin{equation}
  h(\eta )
= {\alpha\over a} \int_{\eta_0}^\eta \mu\sqrt{\gamma} \,
  d\eta + {\alpha\over a} C_i
\end{equation}
where $\eta_0$ is some initial time where the initial conditions
are to be imposed. The constant $C$ entering Eq.~(32) is denoted
$C_i$ at the $i$ stage and will have the labels $e$ and $m$
at the $e$ and $m$ stages.  All (complex) solutions to our
perturbation problem for a given wave vector ${\bf n}$
are completely determined by three arbitrary and independent
(complex) constants. Two of them define a solution to Eq.~(38)
(these constants are $A_1$, $A_2$ when Eq.~(40) is applicable).
The third constant, $C_i$, describes the remaining freedom
in our choice of coordinates.  This remaining freedom is not
a misfortune of the theory. On the contrary, it will later
allow us to join our coordinate system right to the comoving
synchronous coordinate system at the $m$ stage.

One can show by using Eqs.~(24), (12), and (41) (and assuming
$\varphi_0^\prime \neq 0$) that
\begin{equation}
  \varphi_1(\eta ) = {1\over \sqrt{2\kappa}}
\Biggl[ {1\over a}\mu -\sqrt{\gamma}\, h \Biggr] \quad .
\end{equation}
One can also find with the help of Eqs.~(11), (13), and (41) that
\begin{equation}
   {p_1 \over \epsilon_1}
 = {\mu^\prime + \mu
   \Biggl( 2{\alpha^\prime \over \alpha}
   + {1\over 2} {\gamma^\prime \over \gamma} \Biggr)
   - a\sqrt{\gamma}
     \Biggl( \alpha + 2{\alpha^\prime \over\alpha}
           + {\gamma^\prime \over \gamma} \Biggr) h
 \over
   \mu^\prime - \mu
   \Biggl( 4\alpha + {1\over 2} {\gamma^\prime\over \gamma} \Biggr)
   + 3a\sqrt{\gamma} \, \alpha h} \, .
\end{equation}
The quantity $c_l$, where $c_l^2/c^2 = p_1/\epsilon_1$,
plays the role of the velocity of sound for the high-frequency
scalar field oscillations (see also Sec.~VII).

Similarly to what is true for gravitational waves, solutions to
Eq.~(38) are different for the high-frequency and low-frequency
regimes. In the former case,
$n^2 \gg |U(\eta )|$ and $\mu\sim e^{\pm in\eta}$.
In the later case,
$n^2 \ll |U(\eta )|$ and two independent solutions are
$\mu_1 \sim a\sqrt{\gamma}$,
$\mu_2 \sim a\sqrt{\gamma}\int d\eta /(a\sqrt{\gamma})^2$.
The functions $\mu_1$, $\mu_2$ generalize the corresponding
solutions for gravitational waves by replacing $a$ with
$a\sqrt{\gamma}$. Specifically for the scale factors (6),
the solutions $\mu_1$, $\mu_2$ are
$\mu_1\sim \eta^{1+\beta}$ and
$\mu_2\sim \eta^{-\beta}$, in agreement with Eq.~(40).

In the high-frequency regime, the term $\mu^\prime$ dominates
the other terms in Eq.~(44). As one could expect, in this regime,
the velocity of sound is equal to the velocity of light,
$c_l^2=c^2$. In the low-frequency regime, that is when
a given mode enters the under-barrier region, the dominant
solution is $\mu_1$ (for a review, see [19]).
Using this solution in Eq.~(44) one can show that,
in this regime, $p_1/\epsilon_1 \approx q(\beta )$,
that is the ``velocity of sound'' is the same as the one
defined by $p_0^\prime /\epsilon_0^\prime$.
In particular, $p_1/\epsilon_1$ goes to $-$1 for
the low-frequency scalar field solutions at the De Sitter
stage, $\beta = -2$.

Equations (31), (35), and (38) have been derived under the condition
$\varphi_0^\prime \neq 0$. However, the final formula (42) gives
the correct result\protect\\
$h(\eta )=C_i/l_0={\rm const}$
in the De Sitter limit $\gamma = 0$, $\varphi_0^\prime =0$.
One can analyze this case separately, referring to the starting
Eqs.~(23)-(26). One can see that Eqs.~(24) and (12) give
$\xi^\prime =0$, $h={\rm const}$.
Equations~(13), (25), and (23) give
$p_1 = 0$, $V_{,\varphi} = 0$, $\epsilon_1 = 0$.
Finally, Eq.~(11) requires
$h_l^\prime = -\eta n^2 h=-\eta n^2C_i/l_0$.
But this solution for $h$, $h_l^\prime$ is precisely solution (33)
which can be eliminated by a coordinate transformation.
In the De Sitter case governed by the scalar field (1) there
is no density perturbations at all.  Note that the function
$\varphi_1(\eta )$, Eq.~(22), remains arbitrary and the wavelengths
of these fluctuations are growing in the course of expansion.
If one wishes, one can attach to $\varphi_1$ such words as
``inflation is pushing the waves beyond the De Sitter horizon''.
Nevertheless, the result will be zero, as long as $\varphi_1$
is not accompanied by perturbations of the gravitational field.
This is an instructive example in order to realize that
to ``stretch the waves outside the causal horizon'' is not
all we need for generation of density perturbations (likewise,
it is not sufficient to simply stretch the electromagnetic
waves ``beyond the horizon'' in order to generate photons).

In order to receive further insight into the statements
discussed on previous pages, the reader of these School
Proceedings may be interested in performing some exercises.
In particular, one will be able to evaluate certain claims
which exist in the inflationary literature and which gradually
became ``standard''.

It should be clear from the previous discussion that
the essence of the parametric (superadiabatic) amplification
mechanism is the interaction of the perturbations with
the variable gravitational field of the FLRW universes.
It was emphasized long ago~[3] that the condition of equality
of the wavelength of the perturbation to the Hubble radius is,
at most, a necessary condition for amplification, but not
a sufficient condition. [This condition is not even a
necessary condition if the velocity of sound of the density
or rotational perturbations (in contrast to gravitational waves)
is much smaller than the velocity of light. In this case,
the amplification may take place long before the wavelength
of the perturbation becomes comparable with the Hubble radius.
What is a necessary condition for parametric amplification
is that the frequency of the pump field should be comparable
with the frequency of oscillations, not the wavelength
of oscillations be comparable with the Hubble radius.]\quad Even
if the wavelength of a given mode has become comparable with
the Hubble radius and the frequency of oscillations has become
comparable with the frequency of the pump field, it does not
mean that the amplification of the mode must take place.
The wavelength of a given mode is always proportional to
the scale factor $a(\eta )$, and the Hubble radius is always
proportional to $a^2/a^\prime$. These are simply their definitions
and there is no indication of any interaction in these kinematical
facts. What we are really interested in is how the amplitude,
not the wavelength, of the mode behaves. We need to know whether
or not there is a coupling of the mode to the pump field,
and what is the explicit form of the coupling (or, equivalently,
what is the effective potential $U(\eta )$ in the dynamical equations
such as Eq.~(38) or Eq.(39)). The wavelength and the Hubble
radius may ``cross'' each other in course of their evolution
as many times as one wants, it does not guarantee any superadiabatic
amplification, if there is no appropriate coupling.
However, in the inflationary literature the reason
for amplification is usually seen in that the inflation
``stretches'' the waves and acts as a ``magnifying lense''
or as a ``microscope''. The explanation of the generating
mechanism goes usually along the lines:\quad ``oscillations
inside the horizon'', ``stretching'',
``crossing outside the horizon'', ``freezing'',
``crossing back'' and~...~the created perturbations seem
to be ready for comparison with observations.
In order to appreciate the difference between the kinematics
and dynamics one can make the following:

\smallskip

\underbar{\it Exercise 1.}

Take your favorite inflationary model, whichever you like.
Consider two test waves with the same wavelength:\quad one
is gravitational wave, another is electromagnetic wave.
Show that their wavelengths behave in exactly the same manner
(``crossing, stretching, crossing back'').
Show that the gravitational wave will be superadiabatically
amplified (and, hence, can be generated quantum-mechanically)
but the electromagnetic wave will not (the corresponding
potential in the dynamical equation analogous to Eq.~(39)
is identically zero).
\bigskip

Let us now turn specifically to density perturbations in
models governed by the scalar field (1). The archetype
inflationary model is the exponential, or De Sitter,
expansion which corresponds to the case
$\varphi_o^\prime = 0$, $\gamma =0$.  This is the model which generates
the scale-invariant spectrum of cosmological perturbations.
Here and in the rest of the paper we will be interested
in what theory says about the numerical values of the
amplitudes that form the scale-invariant (or almost
scale-invariant) spectrum of density perturbations.
It was argued in the text above that, despite
the presence of the nonvanishing scalar field perturbations,
there is no density perturbations at the De Sitter stage.
The scalar field perturbations may oscillate and then
``cross the horizon'' but there will be no accompanying
density perturbations neither ``inside'' nor ``outside''
the Hubble radius. In other words, in the De Sitter case,
the solution to the linearized Einstein equations is a purely
coordinate solution totally removable by an appropriate coordinate
transformation. For models that slightly deviate from the exponential
expansion (the quasi-De Sitter expansion) the function
$\gamma (\eta )$ is no longer strictly zero and some amount
of density perturbations is present. However, if $\gamma$
is a small number the amplitude of density perturbations
is also small, according to Eq.~(42), regardless of whether
the wavelength is shorter or longer than the Hubble radius.
Now, a considerable part of papers on the subject of perturbations
deals with the so-called gauge-invariant potentials~[16].
In order to make contact with other studies one can perform:

\smallskip

\underbar{\it Exercise 2.}

Show that the gauge-invariant potentials for density perturbations
are strictly zero in the De Sitter case ($\gamma =0$) and they
are small if $\gamma$ is a small number, $\gamma \ll 1$.

\bigskip

The next Exercise will be especially instructive in view
of the previous ones.  I believe that the lack of clear
physical picture of the phenomenon may lead to the loss
of immunity to wrong conclusions. It follows from {\it Exercise 2}
that the amplitude of density perturbations was very small if
the expansion rate was very close to the De Sitter expansion
at the time when the wavelength of our interest was first
``crossing outside'' the Hubble radius. Let the wavelength
of our interest be such that it ``crosses inside'' the Hubble
radius today. What is the amplitude of density perturbations
in this mode today?\quad The ``standard'' inflationary formula
says that
\[
    {\delta\rho \over \rho} \Bigg|_H
    \sim {H^2 \over \dot\phi (t_1)}
\]
where $\dot\phi$ is supposed to be taken at the time
$(t_1)$ when the wavelength of the perturbation ``crossed outside''
the Hubble radius during the De Sitter or quasi-De Sitter epoch
(see also Introduction to this paper). In other words, according
to this formula, the predicted amplitudes of the scale-invariant
spectrum go to infinity.

There exist many ``intuitive'' derivations of the ``standard''
formula. They can only illustrate the fact that the physical
intuition of different people is different.  Apparently,
it is perfectly in accord with the intuition of some theorists
that the amplitude which was almost zero initially has transformed
by today into almost infinity, simply because the wavelength
of the perturbation was first crossing the Hubble radius when
the expansion rate was almost the De Sitter one. Instead of
questioning the validity of this formula, theorists went on
deriving the far-reaching consequences of it. This formula
is essentially the basis for discarding certain scalar field
potentials on the grounds that they generate ``too much''
of density perturbations and for claims that the contribution
of gravitational waves to the quadrupole anisotropy is
``negligibly small'' in the limit of the De Sitter expansion.
For instance, the authors of the recent paper~[28] say that
``In the limit that $n_s \rightarrow 1$ ($p \rightarrow \infty$)
gravitational waves do not contribute to COBE's signal,
and the power spectrum for zeta is the flat, Zel'dovich
spectrum''. Obviously, there is nothing wrong with
the generation of gravitational waves in the De Sitter limit
($p\rightarrow \infty$). The reason for the above conclusion
is the authors' formula (5.15) (the analogue of the ``standard''
formula) which states that the amplitudes of density perturbations
today are arbitrarily large in the De Sitter limit
(the denominator goes to zero). One can hear sometimes
that it would be a disaster if the gravitational wave
contribution to the large-angular-scale anisotropy was
dominant since there would be no scalar perturbations
to seed galaxies. I, personally, think that it would
be a disaster to build serious conclusions on unreliable
theoretical formulas.

Since the intuition of different people is different,
we will try to see how the ``standard'' formula was derived
from equations. We will discuss two papers~[18,29].
[I don't want to be unfair with respect to these
authors:\quad we will discuss these papers precisely
because they are discussable, they contain the formulation
of the problem, equations and certain logic in calculations.
In my opinion, this cannot be said about many other
papers on this subject.]

The basic equation of Bardeen {\it et al.}~[29] is their
equation (2.23), the basic equation of Mukhanov {\it et al.}~[18]
is their equation (6.49). These equations were obtained
by manipulating with the original perturbed Einstein's
equations in order to express them in terms of the
gauge-invariant potentials.  Both groups of authors introduce
the gauge-invariant quantity zeta and use it for derivation
of the final conclusions. The quantity zeta by Mukhanov {\it et al.},
$\zeta_{\rm M.F.B}$, is defined by their equation (6.66),
the quantity $\zeta$ by Bardeen {\it et al.}, $\zeta_{\rm B.S.T.}$,
is defined by their equation (2.45). Both group of authors neglect
in their equations the term which contains the square of the wave
number and work with the approximate equations. They claim that
in this approximation the quantity $\zeta$ is constant.
Bardeen {\it et al.} say that $\zeta_{\rm B.S.T.}$
``is nearly time independent throughout the period when
$k/SH \ll 1$''.
Mukhanov {\it et al.} say that $\zeta_{\rm M.F.B}$
is conserved ``when considering wavelengths far outside
the Hubble radius for which $\nabla^2\Phi$ can be neglected''.
In particular, Mukhanov {\it et al.}, by using their
conservation law, arrive at their equation (6.67) which
says that that the final amplitude of perturbations
can be arbitrarily large if the initial equation of state
(at the first Hubble radius crossing) was sufficiently
close to the De Sitter case. [There is a misprint in their
formula (6.67):\quad the position of quantities  $w(t_i)$
and $w(t_f)$ should be interchanged.]\quad On the basis
of their formula (6.67) Mukhanov {\it et al.} come to
formula (6.68) which is an analogue of the ``standard''
formula and which is ``in agreement with (6.66)'' and ``with
the results of previous investigations~...''\quad [They
give references to the papers listed as [29,30] in the present paper.]

\smallskip

\underbar{\it Exercise 3.}

Using the definitions of the gauge-invariant potentials
and notations of the present paper show that the basic
equations of Bardeen {\it et al.}, their (2.23),
and of Mukhanov {\it et al.}, their (6.49), are totally
equivalent to the equation
\[
\left[ \mu^{\prime\prime} + \mu
\left[ n^2 - {(a\sqrt\gamma)^{\prime\prime} \over a\sqrt\gamma}
\right]\right]^\prime
      - {(a\sqrt\gamma)^\prime \over a\sqrt\gamma}
\left[  \mu^{\prime\prime} + \mu \left[ n^2
      - {(a\sqrt\gamma )^{\prime\prime} \over a\sqrt\gamma}
\right] \right] = 0 \, .
\]
Show that this equation can be reduced to the form
\[
  {1\over a^2\gamma}
\left[ a^2\gamma\left( {\mu\over a\sqrt\gamma}\right)^\prime\right]^\prime
   + n^2{\mu\over a\sqrt\gamma}=X \hskip 4.7cm{\rm{(38^\prime )}}
\]
where $X$ is an arbitrary constant. Thus, the basic equations
of Bardeen {\it et al.} and Mukhanov {\it et al.} are totally
equivalent to equation (38$^\prime$). Now, using the definitions
of the zeta functions show that
\[
       \zeta_{\rm M.F.B.} = {1\over 2n^2}\, {1\over a^2\gamma}
\left[ a^2\gamma
\left( {\mu\over a\sqrt\gamma} \right)^\prime \right]^\prime
\]
and
\[
       \zeta_{\rm B.S.T.} = {1\over 2n^2}\, {1\over a^2\gamma}
\left[ a^2\gamma
\left( {\mu\over a\sqrt\gamma} \right)^\prime \right]^\prime
     + {1\over 6\alpha} \left( {\mu \over a\sqrt\gamma}\right)^\prime \, ,
\]
so that $\zeta_{\rm M.F.B.}$ and $\zeta_{\rm B.S.T.}$
coincide in the limit $n^2\rightarrow 0$.
Show that the approximate equation which Bardeen {\it et al.}
and Mukhanov {\it et al.} work with corresponds to the omitting
of the term with $n^2$ in equation (38$^\prime$). Show that
equation (38$^\prime$) gives in this approximation
\[
  \zeta_{\rm M.F.B.} = {X\over 2n^2}
\]
which is a constant indeed. And, finally, compare the exact
equation (38$^\prime$) with the exact Eq.~(38) of the present
paper to find out that the constant $X$ must be equal to zero.
Thus, in the considered approximation, the conservation law
$\zeta (t_i)=\zeta (t_f)$ translates into the statement
$0=0$, and the claimed ``standard'' formula does not follow
from it.

\bigskip

One additional point:\quad note that the condition of applicability
of the approximate equations (without the term with the square of
the wave number) is
\[
   n^2 \ll \big|{(a\sqrt\gamma )^{\prime\prime} \over a\sqrt\gamma}\big|
\]
(loosely speaking, the wave is under the potential barrier),
not $n^2 \ll \alpha^2$ (the wave is outside the Hubble radius).
This will be important in the next Section.

\section{Late Time Evolution of the Perturbations at the Initial Stage}

The main uncertaintes about the evolution of the very early
Universe refer to the times that preceded the epoch
of the primordial nucleosynthesis.  Whatever was the initial stage,
even if it was governed by the macroscopic energy-momentum tensor
of the collection of superstrings,
it supposedly went over by that epoch into the radiation dominated
stage governed by the scale factor $a(\eta ) = l_o a_e(\eta -\eta_e)$.
The constants $a_e$, $\eta_e$ are to be determined from the continuous
joining of $a(\eta )$ and $a^\prime (\eta )$ at the time
$\eta = \eta_1$ of transition from the $i$ stage to the $e$ stage.
If the $i$ stage is described by the scale factors (6), one derives
\[
   a_e = -(1+\beta ) | \eta_1 |^\beta \, , \quad
   \eta_e = {\beta \over 1+\beta} \eta_1 \quad .
\]

In further applications, we intend to use simple solutions (40),
(42) and to make their appropriate joining with perturbations
at the $e$ stage.  However, the function $\gamma (\eta )$, being
equal to the constant $\gamma =(2\!+\!\beta )/(1\!+\!\beta )$
at the $i$ stage, and to the constant $\gamma = 2$
at the $e$ stage, experiences a finite jump at the transition
point $\eta = \eta_1$. This presented no problem for gravitational
waves, since $\gamma^\prime (\eta )$ did not enter the gravitational
wave potential $U_0(\eta )$. But this becomes important for density
perturbations, since the $U_1(\eta )$ part of the potential acquires
increasingly growing values at the end of the $i$ stage for steeper
and steeper transitions of $\gamma (\eta )$.

To deal with the problem, we introduce a parameterized set of
smooth functions $\gamma(\eta )$ that approximate the step
function in the limit of the parameter $\epsilon$ going to infinity:
\begin{equation}
   \gamma (\eta )
 = {4+3\beta \over 2(1+\beta )} + {\beta \over 2(1+\beta )}
   \tanh [\epsilon (\eta - \eta_1 )] \quad .
\end{equation}
For large negative values of $\eta$ the function (45) goes to
$(2+\beta )/(1+\beta )$, and for large positive values of $\eta$
it goes to 2. We may surround the transition time $\eta = \eta_1$
by a thin ``sandwich'' with boundaries at $\eta_1-\sigma$
and $\eta_1+\sigma$. The asymptotic values of $\gamma(\eta )$
are already reached with arbitrary accuracy at the boundaries,
if $\epsilon$ is sufficiently large, $\epsilon \gg 1/\sigma$.

The function (45) can be integrated, see Eq.~(4), to produce the
function $\alpha (\eta )$:
\begin{equation}
  {1\over \alpha (\eta )}
= {\eta \over 1+\beta} + {\beta \over 2(1+\beta )}
\Biggl[ \eta - \eta_1 + {1\over \epsilon} \ln \,
       {e^{\epsilon (\eta -\eta_1)} + e^{-\epsilon (\eta -\eta_1)}
        \over 2} \Biggr] \, .
\end{equation}
Again, for sufficiently large $\epsilon$, the function
$\alpha (\eta )$ quickly approximates\protect\\
$(1+\beta )/\eta$
to the left of the transition point, and
$1/(\eta -\eta_e)$ to the right of the transition point.
These are the values of $\alpha (\eta )$ that are appropriate
for the $i$ stage (6) and the $e$ stage, respectively.

The divergent functions $\gamma^\prime$, $\gamma^{\prime 2}$
and $\gamma^{\prime\prime}$ participate in the potential
$U_1(\eta )$ (Eq.~(37)). The function $\gamma^\prime$ grows as
$\epsilon$ at the point $\eta = \eta_1$.
The function $\gamma^{\prime\prime}$ is equal to zero at
$\eta = \eta_1$, but it grows as $\epsilon^2$ slightly
to the left of this point, and it grows as $-\epsilon^2$
slightly to the right of this point. We assume that the
transition to the $e$ stage has completed at
$\eta = \eta_1+\sigma$, and we let $\sigma$ to go to zero.
The function $U_1(\eta )$ compresses and stretches to
the arbitrarily large positive and negative values when
$\sigma$ goes to zero and $\epsilon$ goes to infinity.
In particular, the potential $U(\eta )$ goes through zero
and the condition
$n^2\ll\big| (a\sqrt\gamma )^{\prime\prime}/(a\sqrt\gamma )\big|$
can be violated even if the wavelength is still longer than
the Hubble radius.  Examining Eq.~(38), one can expect that
the value of $\mu^\prime (\eta )$ at $\eta = \eta_1+\sigma$
will be different from the value of $\mu^\prime (\eta )$
at $\eta = \eta_1-\sigma$. The integration of
$\mu^{\prime\prime}$ in the limits from
$\eta_1-\sigma$ to $\eta_1+\sigma$
gives a jump in $\mu^\prime$ which depends on the value
of the integral from the divergent part of the potential
$U_1(\eta )$.  Fortunately, it is not $\mu^\prime (\eta )$
itself, but a particular combination
$(\sqrt{\gamma}/a)[\mu^\prime -\mu (\alpha +\gamma ^\prime /2\gamma )]$
that we will need to know in our further calculations.
This simplifies the analysis. Due to Eqs.~(29) and (41)
this combination is precisely the function $v(\eta )$
introduced in Sec.~III.

In terms of the function $v(\eta )$, where
\begin{equation}
        v = {\sqrt{\gamma} \over a}
\Biggl[ \mu^\prime - \mu
\Biggl( \alpha + {1\over 2} {\gamma^\prime \over \gamma}
\Biggr) \Biggr] = \gamma
\Biggl( {\mu \over a\sqrt{\gamma}} \Biggr)^\prime \quad ,
\end{equation}
the basic equation (38) takes on the form
\begin{equation}
  \bigl( a^2 v\bigr)^\prime = -n^2 a\sqrt{\gamma} \, \mu \quad .
\end{equation}
The integration of this equation over the thin ``sandwich''
shows that\protect\\
$v|_{\eta =\eta_1+0} = v|_{\eta =\eta_1-0}$.
In other words, the function
$\mu^\prime -\mu\bigl( \alpha +{1\over 2}{\gamma^\prime\over\gamma}\bigr)$
taken right at the beginning of the $e$ stage is equal
to the value of this function taken right at the end
of the $i$ stage (6) times the factor
${1\over \sqrt{2}}\, {\sqrt{2+\beta \over 1+\beta}}$.
In addition to the conditions:
\[
\gamma |_{\eta =\eta_1-0}=\sqrt{{2+\beta\over 1+\beta}}\, ,\quad
\gamma|_{\eta =\eta_1+0}=2\, ,\quad
\mu |_{\eta =\eta_1-0}=\mu |_{\eta =\eta_1+0}\, ,
\]
this establishes the rules for going through the ``sandwich''.

The meaning of this procedure is the following. Instead
of following the gradual growth of the function $\gamma (\eta )$
from some initial small value up to the final value $\gamma =2$
(in which case it is difficult to find an exact solution to Eq.~(38))
we concentrate the entire evolution of $\gamma (\eta )$
to a short interval of time. In the limit, we consider a sharp
jump of $\gamma$ from one constant value (which may be zero
as at the De Sitter initial stage) to another constant value,
but we take into account the form of the potential $U(\eta )$
and the change in the initial exact solution which is generated
by this jump.

\section{Density Perturbations in the Perfect Fluid Matter}

We will now consider Eqs.~(11)-(14) at the perfect fluid stages
governed by the energy-momentum tensor (15).
Similarly to the scalar field case, the longitudinal-longitudinal
part of stresses vanishes, $p_l = 0$.
For the easier handling of arbitrary $\epsilon_1$, $p_1$
one can introduce the following notations:
$p_1/\epsilon_1 = c_l^2 /c^2$ and
$p_0^\prime /\epsilon_0^\prime = c_s^2/c^2$, see Eq.~(7).
These definitions are convenient but, generally speaking,
they have only formal meaning, since both
$p_1/\epsilon_1$ and $p_1^\prime /\epsilon_0^\prime$
can be negative. However, in certain regimes, the quantity
$c_l$ is a genuine longitudinal velocity of sound (see Sec.~VII).
Our first intention is to derive the equation for $h(\eta )$,
analogous to Eq.~(31) and valid for arbitrary nonzero
$p_1/\epsilon_1$.  Specific cases
$c_l^2 = {1\over 3} c^2$ and $c_l^2 = 0$
will be considered separately.

In order to derive the equation for $h(\eta )$
one can essentially repeat the steps that have lead to Eq.~(31).
Find $h_l^\prime (\eta )$ from Eq.~(11) and plug it into Eq.~(14).
Use Eq.~(13), the first derivative of this equation,
and the definition of $c_l^2 /c^2$.
As a result, one arrives at the equation
\begin{eqnarray}
      h^{\prime\prime\prime}
& + & h^{\prime\prime} \Biggl[ 3\alpha + \alpha\gamma
      + 3\alpha {c_l^2\over c^2}
      + {(c_l^2)^\prime \over c_l^2} \Biggr]\nonumber\\
& + & h^\prime \Biggl[ n^2 {c_l^2 \over c^2}
      + 4\alpha^2 + 6\alpha^2 {c_l^2\over c^2}
      - 2\alpha {(c_l^2)^\prime \over c_l^2} \Biggr]
      + hn^2\alpha\gamma{c_l^2 \over c^2} = 0 \, .
\end{eqnarray}

Now, introduce the function $u(\eta )$ according to Eq.~(34)
and use it in Eq.~(49). Equation~(49) can be reduced to
\begin{eqnarray}
      u^{\prime\prime}
& + & u^\prime \Biggl[ 3\alpha + 3\alpha{c_l^2 \over c^2}
      - {(c_l^2)^\prime \over c_l^2} \Biggr] \nonumber\\
& + & u \Biggl[ n^2 {c_l^2 \over c^2}
      + \Biggl( \alpha + {\alpha^\prime \over \alpha} \Biggr)
      \Biggl( 3\alpha {c_l^2 \over c^2} -{(c_l^2)^\prime \over c_l^2}
      \Biggr) + {a^{\prime\prime} \over a}
             + 2{\alpha^{\prime\prime} \over \alpha}
      \Biggr] \nonumber\\
& + & h\Biggl\{ 3\alpha^2\gamma {c_s^2\over c^2}
      \Biggl[ 3\alpha
      \Biggl( {c_l^2\over c^2} - {c_s^2\over c^2} \Biggr)
             -{(c_l^2)^\prime\over c_l^2}+{(c_s^2)^\prime\over c_s^2}
      \Biggr]\Biggr\} = 0 \, .
\end{eqnarray}
The last term in this equation vanishes if
\begin{equation}
  3\alpha{c_l^2\over c^2} - {(c_l^2)^\prime \over c_l^2}
= 3\alpha{c_s^2\over c^2} - {(c_s^2)^\prime \over c_s^2} \quad ,
\end{equation}
which integrates to
\[
  {c_s^2\over c_l^2} = 1- {{\rm const} \over a\alpha^2\gamma} \quad .
\]

An assumption which is usually made for perfect fluids is:
\begin{equation}
    c_l^2 = c_s^2 \quad .
\end{equation}
Note that Eq.~(52) is certainly true for matter with the equation
of state $p=q\epsilon$, where $q$ is a constant, but Eq.~(52)
is not true in general, and it is not true for the scalar field
matter (1) (unless one considers the under-barrier region where
Eq.~(52) is true approximately, see Sec.~III).
Due to Eq.~(51) the last term in Eq.~(50) cancels out.
(If we have not assumed (51), the function $h(\eta )=C(\alpha /a)$
would not have been a solution to Eq.~(49).)

We can now introduce the function $\nu(\eta )$ according to
(compare with Eq.~(36)):
\begin{equation}
    u = {\alpha\sqrt{\gamma} \over a} \, c_s\, \nu \quad .
\end{equation}
In terms of $\nu(\eta )$, Eq.~(50) takes on the form
\begin{equation}
       \nu^{\prime\prime} + \nu
\Biggl[ n^2{c_l^2\over c^2} - W(\eta) \Biggr] = 0 \quad ,
\end{equation}
where
\[
W(\eta ) = {a^{\prime\prime} \over a}
         - {\alpha^{\prime\prime}\over \alpha}
         - {(\alpha^2 \sqrt{\gamma} c_s)^{\prime\prime}
            \over \alpha^2\sqrt{\gamma} c_s} \quad .
\]
The potential $W(\eta )$ depends exclusively on the scale factor
and its derivatives. Having a solution $\nu(\eta )$
to this equation and using Eqs.~(53) and (34), one can construct
the function $h(\eta )$ and the rest of perturbations.
Similarly to the scalar field case, all solutions
for perturbations with a given ${\bf n}$ are defined by three
constants one of which describes the remaining coordinate
freedom. These constants are expressible in terms of
the constants given at the preceding $i$ stage through
the joining of the perturbations at the transition time
$\eta = \eta_1$ from the $i$ stage to the perfect fluid stage.

We will now consider Eq.~(54) specifically at the
radiation-dominated stage $p={1\over 3}\epsilon$.
We have $c_l^2=c_s^2 =(1/3)c^2$, $\gamma = 2$, and the scale factor
\begin{equation}
   a(\eta ) = l_oa_e (\eta - \eta_e )
\end{equation}
where the constants $a_e$, $\eta_e$ are to be determined
from the continuous joining of $a(\eta )$ and $a^\prime (\eta )$
at the transition time $\eta = \eta_1$. The potential $W(\eta )$
vanishes. Equation~(54) simplifies to the familiar equation
\begin{equation}
  \nu^{\prime\prime} + {1\over 3} n^2 \nu = 0
\end{equation}
which describes the time-dependent part of sound wave oscillations
in the radiation-dominated fluid. In what follows, we will be
using the general solution to this equation written in the form
\begin{equation}
  \nu = B_1\, e^{-i{n\over\sqrt{3}}(\eta -\eta_e)}
      + B_2\, e^{i{n\over\sqrt{3}}(\eta -\eta_e)}
\end{equation}
where $B_1$, $B_2$  are arbitrary and independent (complex) numbers
for each individual wave vector ${\bf n}$.

The function $h(\eta )$ is determined by a known solution for
$\nu (\eta )$ and a coordinate solution with arbitrary constant
$C_e$:
\begin{equation}
    h(\eta ) = {\alpha \over a}
    \int_{\eta_1}^\eta \nu d\eta + {\alpha \over a} C_e \quad .
\end{equation}
All other functions are expressible in terms of $h(\eta )$.
In particular,
\begin{equation}
    h_l^\prime
  = {1\over \alpha} [3h^{\prime\prime}
  + 9\alpha h^\prime + n^2h ] \quad .
\end{equation}

The general solution (57) is always oscillatory in $\eta$-time.
The perturbed energy density, pressure, and the associated
gravitational field $h(\eta )$, $h_l(\eta )$ oscillate in space
and time as they should do for sound waves. However,
if one considers these oscillations at intervals of time shorter
than their period, they may appear as consisting of ``growing''
and ``decaying'' solutions. In particular, this happens
if one considers the relatively long waves,\protect\\
${n\over \sqrt{3}} (\eta_1-\eta_e) \equiv y_1 \ll 1$,
at their first oscillation since the beginning of the $e$-stage,
that is while the condition
${n\over \sqrt{3}} (\eta -\eta_e) \ll 1$
is satisfied. Under this condition, the function $h(\eta )$,
Eq.~(58), can be approximated as
\begin{equation}
    h(\eta )
  = {\bar C_e \over \bar\eta^2}
  + {\bar B_1 \over \bar\eta} + \bar B_2 + \cdots
\end{equation}
where
\[
\bar\eta = {1\over \sqrt{3}} (\eta -\eta_e)\, , \quad
\bar B_1 = {B_1+B_2 \over \sqrt{3} l_oa_e} \, , \quad
\bar B_2 = {-in (B_1-B_2) \over 2\sqrt{3} l_oa_e} \,
\]
and
\[
   \bar C_e = {1\over 3l_oa_e}
\left\{ C_e + {i\sqrt{3} \over n}
\Biggl[ B_1 \biggl( 1-e^{-iy_1}\biggr)
      - B_2 \biggl( 1-e^{iy_1}\biggr) \Biggr]\right\} \, .
\]

The common practice [15] is to use the coordinate freedom
for elimination of the ``most divergent'' term in Eq.~(60),
which is also the ``most decaying'' term, if one goes forward
in time. This is achieved by such a choice of $C_e$ that
$\bar C_e =0$, and the first term in Eq.~(60) vanishes.
Then, the energy density perturbation $\delta\epsilon /\epsilon_0$
(use the definition
\[
  \delta\epsilon /\epsilon_0={\kappa\epsilon_1\over 3\alpha^2}Q
\]
and calculate $\kappa\epsilon_1$ according to Eq.~(11))
can be approximated as
\begin{equation}
        {\delta\epsilon \over \epsilon_0} =
\Biggl( {1\over 9} n^2 \bar B_1 \bar\eta
      + {1\over 2} n^2 \bar B_2 \bar \eta^2
      + \cdots \Biggr) Q \quad .
\end{equation}
The part of Eqs.~(60) and (61) which depends on the coefficient
$\bar B_1$ is usually called the ``decaying'' solution,
while the part with the coefficient $\bar B_2$ is called
the ``growing'' solution.

Despite the possibility of identifying (quite artificially)
the ``growing'' and ``decaying'' solutions, density perturbations
at the $e$-stage form a collection of traveling sound waves
with arbitrary amplitudes and arbitrary phases, as long as
constants $B_1$, $B_2$ are arbitrary and independent.
One should not think that simply because the sound waves
have spent some time ``beyond the horizon'',
$n\bar\eta \ll 1$,
they would transform into standing waves at later times
of their history after they came ``inside the horizon'',
$n\bar\eta \gg 1$. To illustrate this point,
let us take into account the spatial part of the perturbations
and consider the contribution $h_{\bf n} (\eta ,{\bf x})$
of a given mode ${\bf n}$ to the total field
$h(\eta ,{\bf x}) = \sum_{\bf n} h_{\bf n}(\eta , {\bf x})$.
This contribution can be written as
\begin{eqnarray}
         h_{\bf n}(\eta ,{\bf x})
& \sim & h_{\bf n}\, e^{i{\bf nx}}
       + h_{\bf n}^\ast \, e^{-i{\bf nx}} \nonumber \\
& \sim & \Bigl( B_{1{\bf n}} \, e^{-in\bar\eta}
       - B_{2{\bf n}} \, e^{in\bar\eta} \Bigr) e^{i{\bf nx}}
       + \Bigl( B_{1{\bf n}}^\ast \, e^{in\bar\eta}
       - B_{2{\bf n}}^\ast \, e^{-in\bar\eta}\Bigr)
         e^{-i{\bf nx}} \nonumber\\
& = & 2| B_{1{\bf n}}|\cos(n\bar\eta - {\bf nx}
      - \varphi_{1{\bf n}})
      - 2| B_{2{\bf n}}|\cos(n\bar\eta
      + {\bf nx} + \varphi_{2{\bf n}})\qquad
\end{eqnarray}
where
$B_{1{\bf n}}= |B_{1{\bf n}} |e^{i\varphi_{1{\bf n}}}$,
$B_{2{\bf n}}= |B_{2{\bf n}} |e^{i\varphi_{2{\bf n}}}$.
The last line in Eq.~(62) shows explicitely that, in general,
one is dealing with waves traveling in opposite directions
with arbitrary amplitudes and arbitrary phases. A standing wave
can only be ``generated by hand'', by assuming that the constants
$B_{1{\bf n}}$, $B_{2{\bf n}}$
are strictly related. This happens if one declares that he/she
is only interested in the ``growing'' solution and puts $\bar B_1 =0$.
Then, the complex amplitudes
$B_{1{\bf n}}$, $B_{2{\bf n}}$ become related:\quad
$| B_{2{\bf n}}| = (-1)^{k+1} |B_{1{\bf n}}|$,\protect\\
$\varphi_{2{\bf n}} = \varphi_{1{\bf n}}-k\pi$,
$(k= 0,1,2,\ldots )$, and the last line in Eq.~(62)
can be transformed to
\[
  h_{\bf n}(\eta ,{\bf x}) \sim 4| B_{1{\bf n}} |
  \cos\, n\bar\eta\, \cos ({\bf nx} + \varphi_{1{\bf n}} )
\]
which is a standing wave indeed.

Standing sound waves at the $e$ stage are responsible for so-called\protect\\
Sakharov's oscillations~[20] in the power spectrum of density
perturbations in the present Universe, at the $m$ stage.
As we have shown, standing sound waves cannot originate somehow
automatically at the $e$ stage, simply because of the transition
from the ``growing''/``decaying'' regime to the oscillating
regime (see also~[21]). If one works with classical density
perturbations at the $e$ stage and makes no additional assumptions,
one can say nothing about the necessity of standing waves,
except of postulating this.  The point of this discussion is
that the quantum-mechanical generating mechanism,
which we are considering in this paper, does really create
standing waves. Standing waves arise for gravitational waves,
rotational perturbations, and density perturbations.
The physical reason for this is that the waves (particles)
are generated in correlated pairs with equal and oppositely
directed momenta (the two-mode squeezed vacuum quantum states).
This is true for waves of any wavelength, as soon as conditions
for their generation are satisfied. Technically, as we will
see later, the second term in Eq.~(60) taken at the beginning
of the $e$ stage turns out to be much smaller than the third
term in Eq.~(60), that is $B_1 + B_2 \approx 0$ for $y_1\ll 1$.

We should now discuss a great difference between sound waves
and gravitational waves with regard to their evolution in time.
The velocity of sound waves at the $e$ stage is only $\sqrt 3$
times smaller than the velocity of gravitational waves,
but their amplitudes behave drastically different.
The amplitude of a gravitational wave decays as $a^{-1}$
in course of time. The amplitude of a sound wave,
as one can see from Eq.~(58), decays as $a^{-2}$,
since $\alpha\sim a^{-1}$. This leads to a difference
in solutions even for relatively long waves which did not
complete even one cycle of oscillations during the entire
$e$ stage from $\eta =\eta_1$ to $\eta = \eta_2$.
As an illustration, let us consider sound waves which barely
reached the oscillating regime by the end of the $e$ stage.
Their wave numbers satisfy the condition
${n\over\sqrt{3}}(\eta_2-\eta_e)\equiv {n\over n_c}\equiv y_2\approx 1$.
These are the waves whose wavelength was of the order of
the Hubble radius at the time of transition from the $e$ stage
to the matter dominated $m$ stage.  Assuming that the present
day Hubble radius is $l_H \approx 6\cdot 10^3$~Mpc and that the
present day scale factor
$a(\eta_R)$ is $a(\eta_R )\approx 10^4\, a(\eta_2)$,
their wavelength today
$\lambda_c = 2\pi a(\eta_R)/n_c$
is about 220~Mpc. The usual practice, in addition to eliminating
$\bar C_e$, is to concentrate on the ``growing'' solution,
that is to assume that at the beginning of the $e$ stage the second
term in Eq.~(60) is smaller, or at least not larger, than the third
term.  Under these conditions, the final numerical value of the
function $h(\eta )$ is of the same order of magnitude as the initial
value, $h(\eta_2)\approx h(\eta_1)$, for the wavelengths of our
interest. In other words, if the preceding $i$ stage has produced
$h(\eta )$ with some initial numerical value $h(\eta_1 )$,
this number will effectively be transmitted to the beginning of
the $m$ stage.

However, in the very same coordinate system where the
``most decaying'' term in Eq.~(60) was eliminated,
the time derivative $h^\prime (\eta )$ was left large.
The final value of $h^\prime (\eta )$ for the waves
of our interest is $h^\prime (\eta_2)\approx nh(\eta_2)$.
According to Eq.~(12), the function
$h^\prime (\eta )$ describes velocity of matter.
This velocity will be inheritted by matter at
the matter-dominated stage. But this is not velocity
of the fluid elements with respect to each other,
this is not velocity describing deformations of the medium,
and the functions $h^\prime (\eta )$, $h_l^\prime (\eta )$
are not the ones that we may use for our later calculation
of the varations in CMBR. The function $h^\prime (\eta )$
describes velocity of matter with respect to the coordinate
system that we have chosen for our convenience of eliminating
the ``most decaying'' term. At the $m$ stage, however,
we are more interested in the comoving coordinate system.
We are interested in the density contrasts and deformations
of the fluid itself, we are interested in the components
of the accompanying gravitational field that we may use
for calculations of $\delta T/T$. We need a coordinate
system which provides vanishing of $h^\prime (\eta )$
by the end of the $e$ stage. This requires a different
choice of $C_e$. Under this new choice of $C_e$,
the first term in Eq.~(60) survives, and numerically the same,
as in the previous example, initial value $h(\eta_1)$ transforms
into a small number $h(\eta_2) \approx y_1^2h(\eta_1)$
by the beginning of the $m$ stage. This is what we need
to keep in mind when we will compare the amplitudes
of density perturbations and gravitational waves.

Finally, let us consider density perturbations at the $m$ stage,
$p=0$.  At this stage, one has
$c_l^2 = c_s^2 = 0$, $\gamma = {3\over 2}$, and the scale factor
\begin{equation}
   a(\eta ) = l_oa_m (\eta -\eta_m)^2 \quad .
\end{equation}
The scale factor and its first time-derivative are continuous
at the time $\eta = \eta_2$ of transition from the $e$ stage
to the $m$ stage. Therefore,\protect\\
$a_m = a_e/4(\eta_2-\eta_e)$, $\eta_m=-\eta_2+2\eta_e$.
It follows from Eqs.~(13) and (14) that the general solution
for $h(\eta )$, $h_l(\eta )$ has the following form
\begin{eqnarray}
      h
& = & C_1 + {\alpha\over a} C_m \, , \quad
     h^\prime = -{3\over 2} {\alpha^2\over a}\, C_m \, , \nonumber\\
     h_l^\prime \,
& = & {1\over 5} C_1n^2 (\eta -\eta_m) + {1\over a} n^2C_m
    + C_2 {(\eta_2-\eta_m)^3 \over (\eta - \eta_m)^4} \quad .
\end{eqnarray}
The energy-density and velocity perturbations can be found from
Eqs.~(11) and (12). Similarly to the preceding $i$ and $e$ stages,
the perturbations are completely determined by three constants
$C_1$, $C_2$, $C_m$ one of which, $C_m$, reflects the remaining
coordinate freedom.

The constant $C_m$ is entirely responsible for a possible relative
velocity of our fluid with respect to a chosen synchronous coordinate
system, see Eqs.~(12) and (10). A given coordinate system is not
comoving, $T_o^i \neq 0$, as long as $C_m\neq 0$, $h^\prime \neq 0$.
But we know that for a dust-like fluid without rotation one can
always introduce a coordinate system which is both synchronous
and comoving. This is reflected in our ability to remove the
$C_m$ term from $h(\eta )$ and $h_l(\eta )$ by a coordinate
transformation (33). Thus, the choice $C_m = 0$ in Eq.~(64)
is not a restriction of the physical content of the problem,
it is an allowed choice of the coordinate system.

It is here that we will eventually restrict our coordinate freedom,
we will put
\begin{equation}
    C_m = 0 \quad .
\end{equation}
The constants $C_e$ and $C_i$ will not be arbitrary any longer,
they will be determined from the continuos joining of solutions.
We need the comoving coordinate system for simple and appropriate
formulation of the $\delta T/T$ problem. We are interested in the
temperature of CMBR and its anisotropy seen by a comoving observer,
that is by an observer whose world line is one of the matter's world
lines. One of these idealized comoving observers is an observer
on Earth (up to accuracy of some nonzero peculiar velocity,
local rotation, etc.). We are much less interested in feelings
of an observer who wanders in the Universe with arbitrary
time-dependent velocity. In the comoving coordinate system,
the world line of a comoving observer is described by simple
equations $x^i={\rm const}$.

Upon the choice of $C_m = 0$, the perturbations reduce to
\begin{eqnarray}
   h(\eta ) & = & C_1 \nonumber\\
 h_l(\eta ) & = & {1\over 10} C_1n^2 (\eta - \eta_m)^2 - {1\over 3}
               C_2 {(\eta_2-\eta_m)^3 \over (\eta -\eta_m)^3} \quad .
\end{eqnarray}
{}From Eqs.~(66) and (11) one can derive the familiar expression~[15]
\begin{equation}
  {\delta\epsilon \over \epsilon_0} =
\Biggl[ {1\over 20} C_1n^2 (\eta -\eta_m)^2
       -{1\over 6} C_2 {(\eta_2-\eta_m)^3\over (\eta -\eta_m)^3}
\Biggr] Q \, .
\end{equation}
(One may wish to correct a misprint in Eq.~(115.21) of
Ref.~[15]:\quad the decaying solution behaves as
$\eta^{-3}$, not as the printed  $\eta^{-2}$.)\quad As long
as the constants $C_1$, $C_2$ are arbitrary, the power spectrum
of the density perturbations is arbitrary. In particular,
there is no Sakharov's oscillations, {\it a priori},
and they do not arise simply because of the transition from
the $e$ stage to the $m$ stage. For instance, one can start
from a perfectly smooth spectrum at $\eta =\eta_2$ and extrapolate
these data back in time up to the beginning of the $e$ stage.

For the further calculations of $\delta T/T$ we will need
the first time derivatives of the gravitational field
perturbations at the $m$ stage in the comoving coordinates.
Since for the scalar component of the perturbations one has
$h^\prime = 0$, it is only the longitudinal-longitudinal component
$h^\prime_l$ that is effective.

\section{Joining the Perturbations at the Three Stages}

We are now in the position to start our operation of
joining the solutions at $i$, $e$, and $m$ stages.
We want to derive from the first principles the expected
density perturbations at the $m$ stage. Of course,
the result will depend on the unknown behaviour of $a(\eta )$
at the $i$ stage. But this is precisely why we are doing
this study:\quad we try to learn something about the evolution
of the very early Universe by deriving the expected variations
in CMBR and comparing them with the observations.

The general rule for joining solutions to Einstein's equations
is to match from the both sides the intrinsic and extrinsic
curvatures of the transition hypersurface~[22]. For our
solutions written in the class of synchronous coordinate systems,
this translates into the continuity of the spatial metric
and its first time derivative. Since we have already assumed
that $a(\eta )$ and $a^\prime (\eta)$ join continuously,
it is the continuous joining of
$h(\eta )$, $h_l(\eta )$, $h^\prime (\eta )$, and $h_l^\prime (\eta)$
that should be ensured. In fact, it is sufficient to follow
$h(\eta )$, $h^\prime (\eta )$, and $h_l^\prime (\eta)$
as $h_l(\eta )$ is derivable from $h_l^\prime (\eta )$
at all three stages up to the integration constant which can
be removed by the remaining integration constant in Eq.~(33)
anyway. It is convenient to write $h_l^\prime$ at the
$i$ and $e$ stages, respectively, in the form (use Eqs.~(29),
(59) and (42), (58)):
\begin{eqnarray}
      h_l^\prime
& = & {n^2 \over \alpha} h + {\sqrt{\gamma} \over a}
      \Biggl[ \mu^\prime - \mu \Biggl( \alpha + {1\over 2}
      {\gamma^\prime \over \gamma} \Biggr) \Biggl] \nonumber\\
      h_l^\prime
& = & {n^2\over\alpha}h +{3\over a}(\nu^\prime -\alpha\nu) \quad .
\end{eqnarray}
We will denote
$a(\eta )$, $\alpha (\eta )$ at $\eta = \eta_1$ and
$\eta = \eta_2$ by $a_1$, $\alpha_1$ and $a_2$, $\alpha_2$
respectively. It is also convenient to introduce the parameter
$y= {n\over \sqrt{3}} (\eta -\eta_e)$ and its values
$y_1$, $y_2$ at the transition points.

We will first make the joining of solutions in general form,
without adopting any particular coordinate system,
any particular behavior at the $i$ stage, and any particular
wavelength of the perturbations.

Let us start from the $i$-$e$ transition. Whatever was
the $i$ stage and its late time behavior, it produced
certain $\mu (\eta_1)$, $\mu^\prime (\eta_1)$ and ended
at $\eta = \eta_1$ with some values of the scale factor
and its derivatives. For generality, we do not assume,
for the time being, that $\gamma (\eta_1 )$ is exactly 2
and $\gamma^\prime (\eta_1)$ is exactly zero.

{}From the continuous joining of
$h(\eta )$, $h^\prime (\eta )$, and $h_l^\prime (\eta )$
one derives:
\begin{equation}
   C_e
= \int_{\eta_0}^{\eta_1} \mu \, \sqrt{\gamma} \, d\eta + C_i \quad ,
\end{equation}
\begin{equation}
     B_1\, e^{-iy_1} + B_2\, e^{iy_1} \,
  = \sqrt{\gamma} \, \mu + \alpha (2-\gamma )
     \Biggl[ \int_{\eta_0}^{\eta_1} \mu \, \sqrt{\gamma} \,
              d\eta + C_i \Biggr] \, ,
\end{equation}
\begin{equation}
     B_1\, e^{-iy_1} (1+iy_1) + B_2\, e^{iy_1}(1-iy_1)
   = -{1\over 3} a\, \sqrt{\gamma} \,
     \Biggl[ \mu^\prime -\mu
     \Biggl( \alpha + {1\over 2} {\gamma^\prime \over \gamma}
     \Biggr) \Biggl] \, .
\end{equation}
All functions in these equations are taken at $\eta = \eta_1$
so that, for instance, $\gamma$ means $\gamma (\eta_1)$,
$\mu^\prime$ means $\mu^\prime (\eta_1)$, $\gamma^\prime$ means
$\gamma^\prime (\eta_1)$, etc.  For a more compact record we
will also use the following notations:
\[
  u_1 = {\alpha_1\over a_1} \,
        \sqrt{\gamma (\eta_1)} \, \mu (\eta_1) \, , \quad
  v_1 = {\sqrt{\gamma (\eta_1)}\over a_1}
        \Biggl[ \mu^\prime (\eta_1)-\mu (\eta_1)
        \Biggl( \alpha_1 + {1\over 2} {\gamma^\prime\over \gamma}
                (\eta_1 ) \Biggr) \Biggr] \, ,
\]
and
\[
   \bar C_i = C_i + \int_{\eta_0}^{\eta_1}
         \mu\, \sqrt{\gamma}\, d\eta \quad .
\]
Equations (69)-(71) allow us to express the constants
$B_1$, $B_2$, and $C_e$ describing the perturbations at the
$e$ stage entirely in terms of the output values of the functions
defined at the $i$ stage.

Let us now turn to the $e$-$m$ transition. Again, from
the joining of $h(\eta )$, $h^\prime (\eta )$, and
$h_l^\prime (\eta )$ one derives
\begin{equation}
C_m = {4\over 3} C_e - {2\over \sqrt{3}\, n}
      \biggl( B_1\, e^{-iy_1} s + B_2\, e^{iy_1} s^\ast \biggr) \, ,
\end{equation}
\begin{equation}
      C_1 = -{\alpha_2 \over 3a_2} C_e + {1 \over 6a_2y_2}
     \bigl[ B_1\, e^{-iy_1}(s+3y_2 \, e^{-id} ) + B_2\, e^{iy_1}
     (s^\ast + 3y_2 e^{id}) \bigr] \, ,
\end{equation}
\begin{eqnarray}
       C_2 =
 & - & {2\sqrt{3}\, ny_2 \over 5a_2} C_e + {6\over 5a_2}
      \Bigl\{ B_1\, e^{-iy_1}
      \bigl[ e^{-id}(-5+2y_2^2-6iy_2) + iy_2 \bigr] \nonumber\\
 & + & B_2\, e^{iy_1} \bigl[ e^{id} (-5 + 2y_2^2 + 6iy_2)
     - iy_2 \bigr] \Bigr\} \, ,
\end{eqnarray}
where $d=y_2 -y_1 = {n\over \sqrt{3}} (\eta_2-\eta_1)$,
$s=2i+(y_2-2i)e^{-id}$.
Everything at the $m$ stage is known as soon as the coefficients
$C_1$, $C_2$, and $C_m$ are known. They are expressed in terms
of the  coefficients  $B_1$, $B_2$, and $C_e$ attributed to the
$e$ stage.  Since these numbers, in turn, are known implicitly
in terms of the coefficients attributed to the $i$ stage,
we have linked the very beginning with the very end.

Our next step is to impose the requirement (65) and to choose
the comoving coordinate system at the $m$ stage.
{}From Eqs.~(72) and (69) one can find
\begin{equation}
  \bar C_i D = {i\over 2n^2} y_1^2a_1
\Bigl\{ s\bigl[ 3u_1(1-iy_1) + v_1 \bigr]
       - s^\ast\bigl[ 3u_1(1+iy_1) + v_1 \bigr] \Bigr\} \, ,
\end{equation}
where
\begin{eqnarray}
      D
& = & 2y_1^2 + i{2-\gamma \over 2}
      \bigl[ s^\ast(1+iy_1)-s(1-iy_1)\bigr] \nonumber\\
& = & 2y_1^2 + (2-\gamma)
      \bigl[ 2-(2+y_1y_2)\cos\, d-(y_2-2y_1)\sin\, d\bigr] \, .
\end{eqnarray}
Equation (75) says how to choose $C_i$ at the $i$ stage in order
to match right to the comoving coordinate system at the $m$ stage.
We can now put Eq.~(75) into Eq.~(70) and solve Eqs.~(70) and (71):
\begin{equation}
      B_1\, e^{-iy_1} \,
   = {i\over 6D} {a_1\over\alpha_1}
     \{ 6y_1(1-iy_1)u_1 - [(2-\gamma )s^\ast -2y_1]v_1\} \, ,
\end{equation}
\begin{equation}
     B_2\, e^{iy_1}
   = -{i\over 6D} {a_1\over\alpha_1}
     \{ 6y_1(1+iy_1)u_1 - [(2-\gamma )s-2y_1]v_1\} \, .
\end{equation}
Substituting these formulae and Eq.~(75) into Eqs.~(73) and (74),
we reach our goal~---~the finding of $C_1$, and $C_2$
in the comoving coordinates:
\begin{equation}
C_1 = {a_1 \over 3a_2\alpha_1 D}
      \{ 3u_1y_1(\sin \, d+y_1\, \cos\, d)
         - v_1 [(2-\gamma )(\cos\, d-1) -y_1\, \sin\,d]
      \} ,
\end{equation}
\begin{eqnarray}
      C_2 =
& - &  {2a_1 \over 5a_2\alpha_1D}
      \bigl\{ 3u_1y_1
      \bigl[ (10-3y_2^2-10y_1y_2) \sin \, d \nonumber\\
& + &  (-10y_2+10y_1-3y_1y_2^2) \cos\, d \bigr] \nonumber\\
& - &  v_1 \bigl\{ -2(2-\gamma )(5+y_2^2)
      -\bigl[ y_1(10-3y_2^2) - 10y_2(2-\gamma )\bigr]
        \sin\, d \nonumber\\
& + & \bigl[ (2-\gamma )(10-3y_2^2) + 10y_1y_2 \bigr]
      \cos\, d \bigr\} \bigr\} \, .
\end{eqnarray}
So far, no approximations have been made. We will start making them now.

The numerical value of the denominator D, and hence the absolute values
of $C_1$ and $C_2$, depend critically on how close the exiting value\protect\\
$\gamma (\eta_1)\equiv \gamma_1$ is to 2. We are interested
in wavelengths that were longer than the Hubble radius at
$\eta = \eta_1$. Their wave numbers satisfy the requirement
$y_1 \ll 1$.  On the other hand,
$y_2/y_1 = a_2/a_1 \gg 1$, and $d = y_2-y_1\approx y_2$.
The wavelenghts longer than the present day
$\lambda_c = 2\pi a(\eta_R)/n_c \approx 220$~Mpc
correspond to small $y_2$, and $n < n_c$.
These are the wavelenghts of the major interest for
the discussion of the large-angular-scale anisotropy in CMBR.
In the approximation of small $y_2$, two leading terms in
$D$ are
\[
  D \approx\gamma_1y_1^2 + {2-\gamma_1\over 12}y_2^4 \quad .
\]
The second term is much larger than the first one (and, hence,
the expected $C_1$, and $C_2$ are hopelessly small) for all
\[
   n_c{a_1 \over a_2} \sqrt{12\gamma_1/(2-\gamma_1)} < n < n_c \quad ,
\]
unless the exiting value $\gamma_1$ is so close to 2 that
the second term can be neglected.  In order to deal with the most
favorable situation and not proliferate complications,
we will assume that $\gamma_1=2$ and
$(\gamma^\prime /\gamma )(\eta_1)=0$.
We know, see Sec.~IV, that the transition from the very end
of the $i$ stage (after a thin ``sandwich'' interval)
to the very beginning of the $e$ stage can be made arbitrarily
smooth (at least, in theory). So, we will be using
$D \approx 2y_1^2 \ll 1$.

We should now take into account the fact that $v_1 \ll u_1$
for all wavelengths of our interest. Indeed, we are interested
in modes that have interacted with the potential barrier in
Eq.~(38) and have been amplified at the $i$ stage.
For the scale factors (6), their wave numbers satisfy the
requirement $(n\eta_1)^2 \ll \beta (\beta +1)$ which translates
into the condition $y_1^2 \ll \beta /3(\beta +1)$, or simply
$y_1^2 \ll 1$. We will derive the approximate formulae valid
in the leading order by the parameter $y_1$.

As we know, there are two independent solutions in the
under-barrier region. Which of them dominates is the matter
of choice of the initial conditions at $\eta = \eta_o$
(choice of constants $A_1$, and $A_2$ in Eq.~(40)).
For classical solutions, one can choose the initial data
in such a way that there will be no amplification at all,
or there will be even attenuation instead of amplification.
However, a ``typical'' choice of initial data at $\eta = \eta_o$,
which amounts to the averaging over the initial phase (or a rigorous
quantum-mechanical treatment), always leads to the dominant solution
$\mu \sim a\sqrt{\gamma}$, and to amplification~[19].  This is
the choice that we imply here and will justify later, Eq.~(102)
in Sec.~VIII. Since  $v\sim (\mu /a\sqrt{\gamma})^\prime$ and
$\mu \sim a\sqrt{\gamma}$, the quantity $v_1$ is relatively
small. Concretely, for solutions (40), one has
\begin{equation}
        \mu (\eta_1)
\approx {A_1 \over 2^{\beta +{1\over 2}} \,
        \Gamma (\beta +{3\over 2})} (n\eta_1)^{\beta + 1}
\end{equation}
and
\[
          (\mu^\prime -\mu\alpha )(\eta_1 )
\approx - {nA_1 \over 2^{\beta +{1\over 2}} \,
          \Gamma (\beta +{3\over 2})}
          (2\beta + 3)^{-1} (n\eta_1)^{\beta +2}
\]
so that
\[
          {v_1 \over u_1}
\approx - {3(1+\beta ) \over 2\beta +3} y_1^2 \ll 1 \, .
\]
Thus, we can neglect all the terms containing $v_1$ in Eqs.~(77)-(80).

In the leading order, one has
\[
    B_1 \approx - B_2 \approx {i\over 2y_1}\,
                  \sqrt 2 \, \mu (\eta_1) \equiv B
\]
\[
    B_1+B_2 \approx By_1^3 \quad .
\]
These formulae ensure the standing wave pattern for all
wavelengths at the $e$ stage and the power spectrum modulation
at the $m$ stage (compare with Sec.~V).  The leading order
expressions for $C_1$, $C_2$ are as follows:
\begin{eqnarray}
            C_1
& \approx & {1\over 2a_2} {1\over y_1} \sqrt 2 \,
            \mu (\eta_1)\sin\, d \\
            C_2
& \approx & -{3\over 5a_2} {1\over y_1} \sqrt 2 \, \mu (\eta_1)
           \biggl[ (10-3y_2^2)\sin\, d - 10y_2\cos \, d \biggr]
\end{eqnarray}

We will give a brief analysis of Eqs.~(82) and (83).
The growing and decaying components of
$h_l(\eta )$ and $\delta\epsilon /\epsilon_0$,
see Eqs.~(66) and (67), are of the same order of magnitude at
$\eta = \eta_2$ for all wavelengths. The coefficients
$C_1$, $C_2$ are smooth for long waves,
$d\approx y_2 \ll 1$, $n \ll n_c$:
\begin{equation}
   C_1 \approx {1\over 2a_1} \sqrt 2 \,  \mu (\eta_1) \, ,\quad
   C_2 \approx -{6\over 5a_1}
   \Biggl( {n\over n_c} \Biggr)^2 \sqrt 2 \, \mu (\eta_1)
\end{equation}
and are oscillating for shorter waves, $n > n_c$ (Sakharov'
s oscillations).  At a series of frequencies, the factor
$\sin \, d$ is zero, and the growing component totally
vanishes:\quad  no gravitational field perturbations,
no time derivatives of the perturbations, no energy
density perturbations. These modes were highly excited,
like others, at the end of the $i$ stage, but were stripped
off of their energy by the very late times of their evolution.
In terms of quantum mechanics, one can say that these modes have
been desqueezed, sent back to the vacuum state~[23].\quad  [The
generating process and associated squeezing can be visualized
as a depletion of a zero-momentum pump ``quantum'' $2\omega$
into a pair of quanta which have the half-frequency $\omega$
and oppositely directed momenta. Then, the desqueezing can
be thought of as the reverse process.]\quad The position
of zeros is determined by
\[
{n\over \sqrt 3} (\eta_2-\eta_1) =\pi k \, ,\qquad
  k =1,2,3\, \ldots
\]
or, approximately, by
$(n/n_c) = \pi k$.  If one defines the distance $x$
traveled by sound waves between the barriers at
$\eta = \eta_1$ and $\eta =\eta_2$ by
$x=a{1\over \sqrt 3} (\eta_2-\eta_1)$,
the zeros arise when $x$ is covered by an integer number
of half-waves,\protect\\
$x ={\lambda\over 2} k$.  The first zero
in the spectrum of the growing component arises at $k=1$
which corresponds to the present day scale of the order
of 70~Mpc.

The numerical values of $C_1$ and $C_2$ as functions of $n$
are controlled by the $n$-dependent function $\mu (\eta_1)$.
For simple scale factors (6) and solutions (40), $\mu (\eta_1)$
is given by Eq.~(81) where the value of $A_1$ is determined
by quantum mechanics, as will be discussed in Sec.~VIII.
The function $\mu (\eta_1)$ is exactly the same as the one
used for gravitational wave calculations~[17]. This allows
us to make certain comparisons of density perturbations with
gravitational waves.  For instance, the growing component
of gravitational waves taken at the beginning of the $m$ stage,
$h_{gw}(\eta_2)$, has the following amplitude in the low
frequency limit, $n\ll n_c$:
\[
  h_{gw} (\eta_2) \approx {3\sqrt {3\pi} \over \sqrt 2}
  {1\over a_1} \mu (\eta_1) \quad .
\]
This number is $3\sqrt {3\pi}$ times larger than $C_1$, Eq.~(84),
which gives the amplitude $h$ for density perturbations in
the same limit. We can also compare the growing component of
$h_l$ with the growing component of gravitational waves.
Let us take $\eta = \eta_R$ and $n\ll n_H$ where
$n_H \equiv 4\pi /(\eta_R-\eta_m)$ corresponds to the wavelength
equal to the Hubble radius at $\eta = \eta_R$. One can derive
\[
         h_l(\eta_R) \approx {8\pi\sqrt\pi \over 15}
\Biggl( {n\over n_H} \Biggr)^2 h_{gw} (\eta_R) \quad .
\]
As we see, gravitational waves and density perturbations
``enter'' the (time-dependent) Hubble radius with approximately
equal amplitudes, regardless of the numerical values of parameters
$l_o$, $\beta$ describing the $i$ stage.

According to our definitions, see Sec.~VIII, the Fourier component
of the quantized field includes the factor $l_{pl}/\sqrt {2n}$
in addition to $h(\eta )$ or $h_l(\eta )$. We can give an estimate
for the ``characteristic'' amplitude $h(n)\sim nl_{pl} C_1$
of the $h$ field, which is a substitute for a more rigorously
defined expectation value of the dispersion (square root of
the variance) of the field per logarithmic frequency interval.
Combining Eqs.~(84) and (81)
we can find in the low frequency limit
$n\ll n_c$:  $h(n)\sim(l_{pl} /l_o) n^{\beta +2}$, {\it i.e.,}
exactly the same behavior as for gravitational waves.
The growing components of $h_l(\eta )$ and
$\delta\epsilon /\epsilon_o(\eta )$ contain the additional
factor $n^2(\eta -\eta_m)^2$ which gives two extra powers
of $n$ in their spectra. There is nothing spectacular about the
De Sitter case $\beta =-2$. The derivative of the Hubble
parameter can be arbitrarily close to zero at the time
when the wavelength of our interest leaves the Hubble radius
at the $i$ stage. The perturbation will have a finite,
not infinite, amplitude today.

\section{Density and Rotational Perturbations\protect\\
in the High-Frequency Limit}

The normalization of the perturbations is determined by quantum
mechanics. We intend to amplify the zero-point quantum fluctuations
of the primeval matter which is, in our case, the scalar field (1).
Before the amplification, the frequencies of the fluctuations
were much higher than the frequency of the gravitational pump
field. To the modes of our interest the surrounding space-time
seemed at the beginning almost flat.  As a preparation
for quantization, we will first consider density and rotational
perturbations in matter placed in the Minkowski space-time,
that is when gravity is totally neglected ($a(\eta )=1$ in Eq.~(2)).

The deformation of an elastic medium is usually described~[24] with
the help of a displacement three-vector $u_i(t,{\bf x})$
which can be written as
\begin{equation}
    u_i = \xi (t) Q_{,i} + \theta (t)Q_i \quad .
\end{equation}
The scalar function $Q$ is defined by Eq.~(8), the vector function
$Q_i$ is defined by the equations
\begin{equation}
    {Q_{i,k}}^{,k} + n^2 Q_i = 0 \, , \quad
    {Q^i}_{,i} = 0  \quad .
\end{equation}
The deformation tensor is
\[
  u_{ik} = {1\over 2} (u_{i,k} + u_{k,i})
         = {1\over 2} \xi (Q_{,i,k} + Q_{,k,i})
         + {1\over 2} \theta (Q_{i,k} + Q_{k,i})
\]
and its trace is $u={u^i}_{,i} = -\xi n^2 Q$.
The stress tensor can be written in the general form
\begin{equation}
  \sigma_{ik} = s(Q_{,i,k} + Q_{,k,i})
              + \rho c_t^2\theta (Q_{i,k} + Q_{k,i})
              + n^2 (2s-\rho c_l^2\xi)Q\delta_{ik}
\end{equation}
where $\rho$ is density, $c_l$ is the longitudinal velocity
of sound, $c_t$ is the transverse (torsional) velocity of sound,
and $s$ is arbitrary function of time. The equations of motion
\[
   \rho{\partial^2u_i \over \partial t^2}
 = {\partial\sigma_i^k \over \partial x^k}
\]
reduce to the oscillatory equations for elastic waves.
In terms of $\eta$-time, they can be written as
\begin{equation}
   \xi^{\prime\prime} + n^2 {c_l^2\over c^2} \xi = 0 \, , \quad
   \theta^{\prime\prime} + n^2 {c_t^2 \over c^2} \theta = 0 \quad .
\end{equation}

We shall now relate the theory of elasticity with the theory of
cosmological perturbations. We shall consider the high-frequency
limit of the perturbations, that is when the scale factor
$a(\eta )$ is almost constant and its variability can be
neglected in comparison with frequencies of the waves.
As we know, the perturbed components of the energy-momentum
tensor for density and rotational perturbations have the general
form (see Section~II and Ref.~[2]):
\[
  T_o^o = -{1\over a^2} \epsilon_1Q \, \qquad
  T_i^o = -T_o^i = {1\over a^2}\xi^\prime Q_{,i}
   + {1\over a^2} \theta^\prime Q_i \, ,
\]
\begin{eqnarray}
  T_i^k & = & {1\over a^2} {p_l\over 2n^2}
            \biggl( {Q_{,i}}^{,k} + {Q^{,k}}_{,i} \biggr)
            - {1\over a^2}\, {\chi\over n^2}
            \biggl( {Q_i}^{,k} + {Q^k}_{,i} \biggr) \nonumber\\
& &\mbox{}+ {1\over a^2} (p_1+p_l){\delta_i}^k Q \, .
\end{eqnarray}
The differential conservation laws ${T_\alpha^\beta}_{,\beta}=0$
can be reduced in the high-frequency limit to the following
equations:\quad the $\alpha =0$ component gives the equation
$-\epsilon_1^\prime + \xi^\prime n^2=0$, which integrates to
\begin{equation}
   n^2\xi = \epsilon_1 \quad ,
\end{equation}
the $\alpha = i$ components give
\begin{equation}
   \xi^{\prime\prime} + p_1 = 0 \, ,\quad
   \theta^{\prime\prime} + \chi = 0 \quad .
\end{equation}
Equations (90) and (91) can, of course, be obtained from
the perturbed Einstein equations as well.

The stress tensor ${\sigma_i}^k$ is connected with the
perturbed components ${T_i}^k$ by
${\sigma_i}^k = -\rho c^2 {T_i}^k$ $(a(\eta )=1$).
{}From the comparison of Eqs.~(87) and (89) one finds
\begin{equation}
   p_1 = n^2{c_l^2 \over c^2}\xi    \, , \quad
  \chi = n^2{c_t^2 \over c^2}\theta \, , \quad
     s = -\rho c^2 {p_l \over 2n^2}  \, .
\end{equation}
With these expressions for $p_1$, $\chi$, Eqs.~(91)
coincide with the wave equations (88). In cosmology
and theory of elasticity, we are dealing essentially
with the same physics.

The quantization of density and rotational perturbations
should be based on Eqs.~(88). Rotational perturbations
have been considered elsewhere~[2]. The quantum-mechanically
generated rotational perturbations can contribute to
the CMBR anisotropy and may be important for the smaller
scale astrophysics. (One should be aware, though,
that there exists also an alternative view on the subject
according to which rotational perturbations are
``irrelevant for cosmology''~[25].)\quad We will
concentrate here on density perturbations. For simple
models of matter, such as perfect fluids and scalar fields,
the functions $\chi$ and $p_l$ vanish. One is left with
the single variable $\xi$ and isotropic stresses.
The quantization of these oscillations in an elastic
material placed in the Minkowski space-time would lead
to the notion of phonons.

There is no wonder that in the case of scalar field matter
the role of $\xi$ is played by $\varphi_1$. If one writes
$\xi (\eta ) = \xi_0 \, e^{-in\eta}$ and
$\varphi_1(\eta ) = \varphi_{10} \, e^{-in\eta}$,
Eq.~(24) gives
\begin{equation}
 in\xi_0 = \varphi_0^\prime \varphi_{10} \quad .
\end{equation}
In the high-frequency limit (large $n$), the first term in
Eqs.~(23) and (25) dominates. With the help of Eq.~(93)
one derives $\epsilon_1=p_1=n^2\xi$.
In other words, for the high-frequency scalar field perturbations,
the velocity of sound is almost equal to the velocity of light.
The perturbations behave as massless scalar particles.

It is the scalar field oscillations that should be normalized by
ascribing a ``half of the quantum`` to each mode. Due to
the Einstein equations the scalar field perturbations are
accompanied by the gravitational field perturbations.
In the high-frequency limit, the second term in Eq.~(43)
can be neglected, the normalization of $\varphi_1$ transfers
to the gravitational field variable $\mu$ and ultimately to $h$
and $h_l$. This is how we will know the initial amplitude for
density perturbations.

\section{Quantization of Density Perturbations}

In the limit of a free massless scalar field placed in the Minkowski
space-time, we have for each mode:
\[
  \delta \varphi_{\bf k}
= \varphi_1Q + \varphi_1^\ast Q^\ast
= \varphi_1(t) \, e^{i{\bf ky}}
+ \varphi_1^\ast (t) \, e^{-i{\bf ky}} \quad .
\]
The total field can be written as
\begin{equation}
  \delta\varphi (t,{\bf y})
= C {1\over (2\pi)^{3/2}} \int_{-\infty}^{\infty} d^3{\bf k}
  {1\over \sqrt {2w_k}}
  \Bigl[ c_{\bf k} \, e^{-iw_kt} \, e^{i{\bf ky}}
       + c_{\bf k}^\dagger \, e^{iw_kt} \, e^{-i{\bf ky}}
  \Bigr] \, .
\end{equation}
The normalization constant $C$ is to be found from the requirement
\[
  \langle 0| \int_{-\infty}^\infty \epsilon d^3{\bf y} |0\rangle
= {1\over 2} \hbar \int_{-\infty}^\infty d^3 {\bf k} w_k
  \langle 0 | c_{\bf k} c_{\bf k}^\dagger
            + c_{\bf k}^\dagger c_{\bf k} | 0\rangle \, ,
\]
where $\epsilon$ is the energy density of the field.
Since in our case
\[
  \epsilon = {1\over 2}
\biggl[ (\delta \varphi_{,0})^2
      + (\delta \varphi_{,1})^2
      + (\delta \varphi_{,2})^2
      + (\delta \varphi_{,3})^2
\biggr] \quad ,
\]
we derive $C = c\sqrt{\hbar}$. Obviously, the normalization
coefficient C includes the Planck constant $\hbar$ but does
not include the gravitational constant.

To write the field operator (94) in the curved space-time (2) one
should make the following replacements:
\[
  {\bf y} = a(\eta ) {\bf x}\, , \quad
  {\bf k} = {1\over a(\eta )} {\bf n} \, ,  \quad
      w_k = {cn \over a(\eta )} \, , \quad
c_{\bf k} = [a(\eta )]^{3/2} c_{\bf n} \, .
\]
The field operator takes on the form
\begin{equation}
  \delta\varphi (\eta ,{\bf x})
= {1\over a(\eta )} \sqrt{c\hbar} {1\over (2\pi )^{3/2}}
  \int_{-\infty}^\infty  d^3{\bf n} {1\over \sqrt{2n}}
  \bigl[ c_{\bf n}\, e^{-in\eta} \, e^{i{\bf nx}}
        + c_{\bf n}^\dagger \, e^{in\eta} \, e^{-i{\bf nx}}
  \bigr]  \, .
\end{equation}
This expression is only valid in the high-frequency limit,
when the field can be regarded as free and its non-adiabatic
interaction with gravity can be neglected. When the interaction
becomes important, the time dependence of the field ceases to be
so simple. The operator $c_{\bf n}e^{-in\eta}$,
should be replaced by $c_{\bf n} (\eta )$,
and the evolution of $c_{\bf n}(\eta )$,
$c_{\bf n}^\dagger (\eta )$, should be found
from the Heisenberg equations of motion.

The matter perturbations are accompanied by the gravitational
field perturbations. Due to the Einstein equations they
are linked together and form, in a sense, a united entity.
As we have seen in Sec.~III, the entire dynamical problem
at the $i$ stage reduces to a single wave equation (38) for
a single variable $\mu (\eta )$. All perturbations
can be found from a given solution to this equation.
It follows from Eq.~(43) that
$\mu (\eta ) \approx \sqrt {2\kappa}\, a\varphi_1(\eta )$
in the high-frequency limit. Since the positive frequency
scalar field ${\bf n}$-mode solution is
$\varphi_1(\eta )\sim (1/a(\eta ))\sqrt{c\hbar}\,c_{\bf n}\,e^{-in\eta}$
this leads to
$\mu (\eta ) \sim 4\sqrt\pi \, l_{pl} c_{\bf n}e^{-in\eta}$.
Note that the normalization coefficient includes the
gravitational constant and is proportional to the Planck
length $l_{pl} =(G\hbar /c^3)^{1/2}$.

Having in mind the basic wave equation (38), we can now introduce
the ``fundamental'' scalar field $\Phi (\eta ,{\bf x})$
which describes the whole quantum system interacting
with the gravitational pump field:
\begin{equation}
        \Phi (\eta , {\bf x})
      = 4\sqrt\pi l_{pl} {1\over (2\pi)^{3/2}}
        \int_{-\infty}^\infty d^3{\bf n} {1\over \sqrt{2n}}
\biggl[ c_{\bf n}(\eta )e^{i{\bf nx}}
      + c_{\bf n}^\dagger (\eta )e^{-i{\bf nx}} \biggr]  \, .
\end{equation}
The annihilation and creation operators
$c_{\bf n}(\eta )$, $c_{\bf n}^\dagger (\eta )$
are governed by the Heisenberg equations of motion
\begin{equation}
  {dc_{\bf n} (\eta ) \over d\eta}
= -i[c_{\bf n}(\eta ), H] \, ,\quad
  {dc_{\bf n}^\dagger (\eta ) \over d\eta}
= -i[c_{\bf n}^\dagger (\eta ), H] \quad .
\end{equation}
The interaction Hamiltonian $H$ is given by
\begin{equation}
H = nc_{\bf n}^\dagger c_{\bf n}
  + nc_{-{\bf n}}^\dagger c_{-{\bf n}}
  + 2\sigma (\eta ) c_{\bf n}^\dagger c_{-{\bf n}}^\dagger
  + 2\sigma^\ast (\eta ) c_{\bf n} c_{-{\bf n}} \, ,
\end{equation}
where the coupling function $\sigma (\eta )$ is
\[
\sigma (\eta )={i\over 2}{(a\sqrt\gamma )^\prime\over a\sqrt\gamma}\, .
\]
Equation~(98) demonstrates explicitly the underlying parametric
interaction of the field oscillators with the pump field.
This is simply a generalization of a theory previously
developed for gravitational waves (for a review, see Ref.~19).

The common way of solving Eqs.~(97) and (98) is to write
the operators in the form
\begin{equation}
  c_{\bf n}(\eta )
= u_n(\eta )c_{\bf n}(0)+v_n(\eta )c_{-{\bf n}}^\dagger (0)\, , \quad
  c_{\bf n}^\dagger (\eta )
= u_n^\ast (\eta ) c_{\bf n}^\dagger (0)
  + v_n^\ast (\eta )c_{-{\bf n}} (0) \, ,
\end{equation}
where $c_{\bf n}(0)$, $c_{\bf n}^\dagger (0)$ are the initial values
of the operators taken at some $\eta = \eta_0$
long before the interaction became effective, and
$[c_{\bf n}(0),c_{\bf m}^\dagger (0)]=\delta^3({\bf n}-{\bf m})$
The classical complex functions
$u_n(\eta)$, $v_n(\eta )$ (do not mix up with the functions
$u(\eta )$, $v(\eta )$ introduced in Sec.~III) obey the condition
$|u_n|^2 - |v_n|^2 = 1$ and satisfy the equations
\begin{equation}
  iu_n^\prime = nu_n + 2\sigma v_n^\ast \, , \quad
  iv_n^\prime = nv_n + 2\sigma u_n^\ast
\end{equation}
with the initial data $u_n(0)=1$, $v_n(0) =0$.
If one introduces
$\mu_n(\eta ) \equiv u_n(\eta )+v_n^\ast (\eta)$,
it follows from Eqs.~(100) that the function
$\mu_n(\eta )$ should satisfy precisely the Eq.~(38).
The initial conditions for $\mu_n(\eta )$ in the
high-frequency limit $|n\eta | \rightarrow \infty$ are
$\mu_n(\eta )\rightarrow e^{-in(\eta -\eta_0)}$,
$\mu_n^\prime (\eta )\rightarrow -in\, e^{-in(\eta -\eta_0)}$.

For each mode ${\bf n}$ there exists the vacuum state
$|0_{\bf n} \rangle$ defined by the condition
$c_{\bf n}(0) | 0_{\bf n} \rangle = 0$. As a result of
the Schr\"odinger evolution with the Hamiltonian (98),
the initial vacuum state
$|0_{{\bf n},-{\bf n}}\rangle\equiv |0_{\bf n}\rangle |0_{-{\bf n}}\rangle$
transforms into a multiparticle two-mode squeezed vacuum state
(see Ref.~19 and references cited therein). In other words,
the perturbations (waves) are generated in correlated pairs.
Squeezing is a phenomenon, not a formalism.

At this point, it may be useful to discuss the difference
between squeezed states and other quantum states.
It is sometimes said that the strongly squeezed states
are ``almost classical'' since their wave functions can be
written in the WKB form and the noncommutativity of the
creation and annihilation operators is not important.
The implication seems to be that there is no need for
quantum-mechanical treatment of cosmological perturbations,
everything can be done ``classically''. I think that these
arguments are misleading. It is true that the strongly
squeezed quantum states are the multiparticle states,
the mean number of quanta in these states is much larger
than one. Long ago, but not now, this was considered as
a criterium for the state to be ``classical''.
However, there are many quantum states which are the
``WKB classical'' and whose occupation numbers are very big
and, yet, the statistical properties of these states can be
entirely different. Two quantum states may have exactly
the same, and big, mean number of quanta but, along with this,
they may have drastically different variances of this number.
This is precisely what is true with regard to squeezed states,
as compared with the ``most classical''
quantum states --- coherent states.
\vfill\eject

\underbar{\it Exercise 4.}

Show that the mean value of the number operator $N=a^\dagger a$
in a squeezed vacuum state is $\langle N\rangle = sh^2r$.
Let the squeeze parameter $r$ be $r\gg 1$.
Consider a coherent state with exactly the same mean number
$\langle N\rangle$.  Show that in the coherent state the variance
$\langle (\Delta N)^2\rangle$
is $\langle (\Delta N)^2\rangle =\langle N\rangle$
while in the squeezed vacuum state it is
$\langle (\Delta N)^2\rangle ={1\over 2}sh^22r\approx\langle N\rangle^2$.

\vskip 1cm

The statistical properties of the field are determined by
$c_{\bf n}(\eta )$, $c_{\bf n}^\dagger (\eta )$.
By using Eq.~(99) one can rewrite Eq.~(96) in the form
\begin{equation}
       \Phi (\eta , {\bf x})
     = 4\sqrt\pi l_{pl} {1\over (2\pi)^{3/2}}
       \int_{-\infty}^\infty d^3{\bf n} {1\over \sqrt {2n}}
\biggl[ c_{\bf n}(0) \mu_n(\eta )e^{i{\bf nx}}
      + c_{\bf n}^\dagger (0 )\mu_n^\ast (\eta ) e^{-i{\bf nx}} \biggr]
\end{equation}
where the functions $\mu_n(\eta )$, $\mu_n^\ast (\eta )$
should be taken with the appropriate initial conditions
discussed above. For simple solutions (40), the initial
conditions translate into the requirements
\begin{equation}
  A_1 = -{i\over \cos\, \beta\pi}
  \sqrt{{\pi\over 2}} \, e^{i(n\eta_0 +{\pi\beta\over 2})} \, ,\quad
  A_2 = iA_1\, e^{-i\pi\beta} \quad .
\end{equation}

The quantized gravitational field perturbations are expressible
entirely in terms of the $\Phi (\eta ,{\bf x})$ field (96).
There are many components of $h_{ij}$ but there is only one sort
of creation and annihilation operators.  Let us introduce new
notations
\[
{\stackrel{1}{h}} (\eta ) = h(\eta )\, ,\qquad
{\stackrel{2}{h}} (\eta ) = h_l(\eta )
\]
and the polarization tensors
\[
{\stackrel{1}{P}}_{ij} = \delta_{ij}\, , \qquad
{\stackrel{2}{P}}_{ij} = -n_in_j/n^2 \, , \qquad {\rm and} \qquad
{\stackrel{1}{P}}_{ij} {\stackrel{2}{P}}^{ij} =-1 \, .
\]
The field operator $h_{ij}(\eta ,{\bf x})$ can be written as
\begin{equation}
  h_{ij}
= 4\sqrt\pi l_{pl} {1\over (2\pi )^{3/2}}
  \int_{-\infty}^\infty d^3{\bf n} {1\over\sqrt{2n}}
  \sum_{s=1}^2
  {\stackrel{s}{P}}_{ij}({\bf n})
  \biggl[ c_{\bf n}(0)
  {\stackrel{s}{h}}_n \, e^{i{\bf nx}} + c_{{\bf n}}^\dagger (0)
  {\stackrel{s}{h}}_n^\ast \, e^{-i{\bf nx}} \biggr]
\end{equation}
where the classical complex functions
${\stackrel{s}{h}}_n(\eta )$ should be derived from
$\mu_n(\eta )$ through the equations and initial conditions
already discussed. A similar expression can be written
for the operator of the energy density perturbation
$\delta\epsilon /\epsilon_o$. Equation~(103) is the starting point
for the calculation of the expected angular anisotropy in CMBR.
\vfill\eject

\section{Variations of the CMBR Temperature Caused\protect\\
by Density Perturbations of Quantum-Mechanical Origin}

The photons of CMBR are emitted at $\eta = \eta_E$ and are received
by us at\protect\\
$\eta = \eta_R$. A particular direction of observations
is characterized by the unit vector
$e^k =(\sin\theta\cos\phi$, $\sin\theta\sin\phi$, $\cos\theta )$.
In absence of perturbations, temperature of CMBR seen in all
directions would be the same, $T$. Gravitational field $h_{ij}$
associated with the density perturbations at the $m$ stage
causes a variation of the temperature with respect to
the unperturbed value $T$~[26]:
\begin{equation}
  {\delta T\over T} (e^k)
= {1\over 2} \int_0^{w_1}
\Biggl( {\partial h_{ij} \over \partial\eta} e^ie^j \Biggr) dw \, ,
\end{equation}
where $w_1=\eta_R-\eta_E$, and $\partial h_{ij}/\partial\eta$
is taken in the comoving synchronous coordinate system along
the path $x^k = e^kw$, $\eta =\eta_R - w$.
In case of small perturbations, which we are actually
dealing with, the emission time $\eta_E$ can be regarded
as being one and the same for all directions.
Since the scale factor satisfies the approximate relationship
$a(\eta_E)/a(\eta_R)\approx 10^{-3}$, and the $\eta$
time can be chosen in such a way that
$\eta_R -\eta_m = 1$, the quantity $w_1$ is close to 1.
We will use $w_1=1$. The wavelength equal to the present
day Hubble radius $l_H$ corresponds to $n_H=4\pi$.

For the quantized $h_{ij}$ perturbations, the temperature variation
$\delta T/T$, Eq.~(104), becomes a quantum-mechanical operator.
Since it is only the longitudinal-longitudinal part of $h_{ij}$
that participates in producing $\delta T/T$, we can write
\begin{eqnarray}
      {\delta T \over T} (e^k)
& = & -2\sqrt\pi l_{pl} {1\over (2\pi )^{3/2}}
      \int_0^1 dw \int_{-\infty}^\infty
     d^3{\bf n} {(n_ie^i)^2\over n^2}\nonumber\\
& \times & \biggl[ c_{\bf n}(0) f_n(\eta_R-w)e^{in_ke^kw}
       + c_{\bf n}^\dagger (0) f_n^\ast (\eta_R-w)e^{-in_ke^kw}
\biggr]\nonumber\\
\end{eqnarray}
where
\[
   f_n(\eta_R-w)= {1\over \sqrt{2n}} {dh_l(\eta)\over d\eta}
\Bigg|_{\eta =\eta_R-w} \quad .
\]

The individual observed distributions of the CMBR temperature
over the sky should be compared with theoretical predictions
based on the quantum-mechanical expectation values.
The mean value of $\delta T/T(e^k)$ is obviously zero,
$\langle 0| \delta T/T\, (e^k)|0\rangle = 0$. The variance
$\langle 0| \delta T/T\, (e^k)\delta T/T(e^k)|0\rangle$
is not zero but does not depend on the point and direction
of observations. To study the angular distribution of the
temperature variations one should construct the angular
correlation function $K$ for two different directions
$e_1^k$ and $e_2^k$,
$e_1^ke_2^i \delta_{ki}=\cos\delta$:
\[
  K = \langle 0|
      {\delta T\over T} (e_1^k)
      {\delta T\over T} (e_2^k) |0\rangle \quad .
\]
By manipulating with the product of two expressions (105),
one can derive
\begin{eqnarray}
  K = & & 4\pi l_{pl}^2 {1\over (2\pi )^3}
          \int_0^1 dw \int_0^1 d\bar w \int_0^\infty n^2 |f_n|^2 \, dn
          \int_0^\pi \sin\theta \, d\theta \nonumber \\
      & & \int_0^{2\pi} {(n_ie_1^i)^2\over n^2} {(n_ie_2^i)^2\over n^2}
          \cos(n_k \zeta^k) d\phi \, ,
\end{eqnarray}
where $\zeta^k = e_1^kw-e_2^k\bar w$.
A lengthy calculation of the integrals over the angular variables
$\phi$, $\theta$ gives the following result:
\begin{eqnarray}
& & \int_0^\pi \sin\theta \, d\theta
    \int_0^{2\pi} {(n_ie_1^i)^2 \over n^2}
    {(n_ie_2^i)^2 \over n^2}\cos (n_k\zeta^k)d\phi\nonumber\\
& = & 4\pi\sqrt{\pi\over 2}\biggl\{ \cos^2\delta (n\zeta )^{-1/2}
    J_{1/2} (n\zeta ) \nonumber \\
& &\mbox{}+ (1\!-\! 5\cos^2\delta )(n\zeta )^{-3/2}
    J_{3/2} (n\zeta) \nonumber\\
& &\mbox{}+ 2\cos\delta(1\! -\!\cos^2\delta ) (nw)(n\bar w)(n\zeta)^{-5/2}
      J_{5/2} (n\zeta) \nonumber\\
& &\mbox{}+ 4(3\cos^2\delta \!-\! 1)(n\zeta )^{-5/2}
    J_{5/2} (n\zeta ) \nonumber\\
& &\mbox{}+ 8\cos\delta (cos^2\delta \! -\! 1)
            (nw)(n\bar w)(n\zeta )^{-7/2}
    J_{7/2} (n\zeta) \nonumber\\
& &\mbox{}+ (cos^2\delta \! -\! 1)^2 (nw)^2(n\bar w)^2
              (n\zeta)^{-9/2}J_{9/2}(n\zeta )\biggr\}
\end{eqnarray}
where
$\zeta = (w^2-2w\bar w \cos\delta + \bar w^2 )^{1/2}$.
For further calculations one may use the following formula
(valid for half-integer $\nu$):
\begin{equation}
  (n\zeta )^{-\nu} J_\nu (n\zeta )
= \sqrt{2\pi} \sum_{k=0}^\infty (\nu + k)
  {J_{\nu +k} (nw) \over (nw)^\nu}
  {J_{\nu +k} (n\bar w) \over (n\bar w)^\nu}
  {d^{\nu-1/2} \over dz^{\nu-1/2}} P_{k+\nu-1/2} (z)
\end{equation}
where $z=\cos\delta$ and $P_l(z)$ are the Legendre polynomials.
(I derived and used this formula in course of studying the
gravitational wave~[17] and rotational~[2] perturbations,
but I believe that this formula may exist somewhere
in the previously published literature.)\quad  With the
help of Eq.~(108), one can rearrange the correlation function
$K$ to the following final expression:
\begin{equation}
   K = l_{pl}^2 \sum_{l=0}^\infty K_l P_l (\cos\delta ) \quad ,
\end{equation}
where
\[
   K_l = (2l+1)\int_0^\infty |I_{ln}|^2 dn \quad ,
\]
\begin{equation}
        I_{ln} = \int_0^n {f_n(\eta_R-x/n) \over x^{1/2}}
\Bigg\{ \Biggl[ {l(l-1) \over x^2} - 1 \Biggr]
        J_{{1\over 2}+l} (x)
     + {2\over x} J_{{3\over 2}+l} (x) \Bigg\} dx \, ,
\end{equation}
and  $x=nw$. The derived formula for $K$ is general and can
be used with arbitrary function $f_n$. In practice,
the limits of integration over $n$ are determined
by the frequency interval within which the perturbations
were really generated.

As we see, the decomposition of $K$ consists of all multipoles
including the monopole, $l=0$, and dipole, $l=1$, terms.
To carry out the calculations up to a concrete number, we will
consider scale factors (6) and solutions (40).

As we know, the growing and decaying components of $h_l(\eta )$
are of the same order of magnitude at $\eta =\eta_2$.
However, the decaying component is decreasing since then
and can be neglected in the calculation of $\delta T/T$.
For the coefficient $C_1$ responsible for the growing solution,
see Eq.~(82), we have
\begin{equation}
   |C_1|^2 \approx {1\over 2l_o^2} | \psi (\beta )|^2
   n^{2\beta} n_c^2 \sin^2 {n\over n_c} \quad ,
\end{equation}
where
\[
        |\psi (\beta )|^2 = {\pi\over 2}
\biggl[ 2^{\beta+{1\over 2}} \cos\beta\pi \,
        \Gamma (\beta + {3\over 2} )\biggr]^{-2} \, , \quad
   |\psi (\beta )|^2 = 1 \quad {\rm for} \, \beta =-2 \quad .
\]
The expression for $K_l$ takes on the form
\begin{equation}
   K_l = {1\over 100l_o^2} |\psi (\beta )|^2 \, n_c^2(2l+1)
\int_0^\infty n^{2\beta +1} \sin^2 {n\over n_c} (i_{ln})^2 dn \, ,
\end{equation}
where
\begin{equation}
  i_{ln} = \int_0^n {n-x \over x^{1/2}}
\Bigg\{ \Biggl[ {l(l-1) \over x^2} -1 \Biggr]
        J_{{1\over 2}+l} (x) + {2\over x} J_{{3\over 2}+l}(x)
\Bigg\} dx \quad .
\end{equation}

We will now estimate the contribution of long waves, $n< 1$,
to the lower order multipoles $K_l$. The integrals $i_{ln}$
should be calculated separately for $l=0$, $l=1$, and $l\geq 2$,
because of the factor $l(l-1)$ in Eq.~(113).  For $l\geq 2$,
the term with this factor dominates. The approximate expression
for $(i_{ln})^2$, $l\geq 2$, takes on the form
\begin{equation}
  (i_{ln})^2 \approx A_l^2 n^{2l} \quad ,
\end{equation}
where
\[
   A_l^2 = \Biggl[ 2^{l+{1\over 2}} \,
  \Gamma \Biggl( l+ {3\over 2} \Biggr) \Biggr]^{-2} \, .
\]
For $l=0$ and $l=1$, the term with the factor $l(l-1)$
does not contribute to $(i_{ln})^2$. The result still
have the form of Eq.~(114) but $n^{2l}$ should be replaced
by ${1\over 100}n^6$ for $l=1$, and by ${1\over 36}n^4$ for
$l=0$.  These results are in full agreement with Ref.~13:
in the limit of long waves, the monopole and dipole
contributions of an individual wave are suppressed;
the monopole component is of the same order of magnitude
as the quadrupole component, while the dipole component is
further suppressed by an extra power of $n$.

We should now use $(i_{ln})^2$ for the calculation of $K_l$.
This is where the spectrum of the perturbations comes into play.
We can write for $l \geq 2$:
\begin{equation}
   K_l = {1\over 100l_o^2} |\psi (\beta )|^2 (2l+1)
         A_l^2 \int_0^1 n^{2\beta +3+2l} dn \quad ,
\end{equation}
and the appropriate replacements discussed above should be made
for $l=1$, $l=0$.  Since we are working in the limit of small $n$,
the integration over $n$ cannot be extended to the values $n>1$.
However, typically, short waves contribute little to
the lower index multipoles. The values of $n$ up to $n\approx 2l$
are more important, but for the purposes of simple evaluation we
restrict the integration by $n = 1$.

The suppression of the monopole contributions of individual
waves saves us from a big trouble. If Eq.~(115) were true
for $l=0$, the monopole term $K_0$ would be power-law divergent
in the limit of $n\rightarrow 0$ for all $\beta < -2$.
In order not to be in conflict with the finite observed 2.7~K
temperature, we would need to resort to the fine tuned minimally
sufficient duration of the $i$ stage, in which case the long waves
with this spectrum are simply not being generated. From the correct
Eq.~(115) it follows that the danger of divergence for $K_0$ and $K_2$
arises only in models with $\beta < -4$. (The quadrupole anisotropy
produced by gravitational waves does also diverge in these
models~[17].)\quad Thus, the interval $-2 \geq \beta > -4$
is potentially allowed.

By the way, the level of the quadrupole anisotropy puts sensitive
qualitative constraints on the conjectures [31] about the
``very inhomogeneous'' Universe on scales ``far beyond the visible
horizon''. Even small linearized perturbations can not have totally
unrestricted amplitudes and spectra on scales longer
than $l_H$, see~[13,32,33].

We will now introduce the notations
$l_{pl}\sqrt{K_0}=M$, $l_{pl}\sqrt{K_1} =D$,\protect\\
$l_{pl}\sqrt{K_2}=Q$, and will compare $M$, $D$, and $Q$.
For the quadrupole $Q$, one can find from Eq.~(115)
\begin{equation}
    Q \approx {1 \over 30\sqrt{5\pi}}
    {l_{pl}\over l_o} |\psi (\beta )|
    {1\over \sqrt{\beta+4}} \quad .
\end{equation}
The monopole $M$ and dipole $D$ are related with $Q$ by
\[
  {M\over Q} = {{\sqrt 5}\over 2} \, ,\quad
  {D \over Q} = \sqrt{{3\over 20}}
  \sqrt{{\beta +4\over \beta + 5}} \quad .
\]
We do not have an observational access to the unperturbed temperature
$T$, but whatever is the measured $Q$, we can expect that
a correction of about the same magnitude as $Q$ is included
in the measured $T$. The same is true for the dipole component
$D$ (there is little doubt, however, that the overwhelming part
of the measured dipole anisotropy is accounted for by our
peculiar motion).

We will now compare the contributions of density perturbations
and gravitational waves to the components  $K_l$ of the correlation
function $K$ in the long wave limit, $n< 1$. We should compare
Eq.~(115) with the analogous expression for gravitational
waves~[17]. The ratio of the gravity wave contribution
${\stackrel{g}{K}}_l$ to the density contribution
${\stackrel{d}{K}}_l$ has the form
\[
        { {\stackrel{g}{K}}_l \over {\stackrel{d}{K}}_l}
\approx {(l+1)(l+2)(2l+1)^2 \over 2l(l-1)} \quad .
\]
We see that the ratio is independent of the parameters
$l_o$, $\beta$ describing the $i$ stage.
For the quadrupoles
${\stackrel{g}{Q}}$ and ${\stackrel{d}{Q}}$,
we have in the long wave limit
\[
     { {\stackrel{g}{Q}} \over {\stackrel{d}{Q}} }
\approx \sqrt{75} \quad ,
\]
that is a somewhat larger contribution of gravitational waves.
It is necessary to take into account also the shorter waves
in order to get a more accurate estimate.

In conclusion, there is no dimensionless ratios that could be
adjusted in such a way that the contribution of density
perturbations to the quadrupole anisotropy would be much larger
than the contribution of gravitational waves.  These
contributions are of the same order of magnitude while numerical
coefficients are somewhat in favour of gravitational waves.  At
the same time, the very generation of density perturbations (and
rotational perturbations) is more problematic than the generation
of gravitational waves.  On these grounds one can conclude that
if the observed large-angular-scale anisotropy of CMBR is caused
by cosmological perturbations of quantum-mechanical origin
(what else?), they are, most likely, gravitational waves.

\section{Acknowledgments}

I appreciate the hospitality of R. Kerner and the Laboratory
of Gravitation and Relativistic Cosmology in Paris, France,
where a part of this paper was written. My thanks also go to
P. Steinhardt, V. Mukhanov, and D. Salopek for critical comments and
correspondence, and to K. Thorne for useful advices.

The work was supported in part by NASA Grant NAGW 2902, 3874
and NSF Grant 92-22902.

\section{References}
\begin{itemize}
\item[1.]
Smoot, G. F., {\it et al.} (1992) {\it Astrophys. J.} {\bf 396}, L1;
Ganger, K., {\it et al.} (1993) {\it Astrophys. J.} {\bf 410}, L57
Proceedings of the Texas/Pascos Symposium,
Berkeley (1992) (special issue) {\it Ann. N.Y. Acad. Sci.} {\bf 688},
\item[2.]
Grishchuk, L. P. (1993) {\it Phys. Rev.} {\bf D 48}, 5581.
\item[3.]
Grishchuk, L. P. (1974) {\it Zh. Eksp. Teor. Fiz.} {\bf 67}, 825;
[{\it Sov. Phys. JETP} {\bf 40}, 409 (1975)];
(1977) {\it Ann. N.Y. Acad. Sci.} {\bf 302}, 439.
\item[4.]
Guth, A. H. (1981) {\it Phys. Rev.} {\bf D 23}, 347;
Linde, A., (1990) {\it Particle Physics and Inflationary Cosmology},
Gordon and Breach, New York;
Kolb, E. W. and Turner, M. S. (1990) {\it The Early Universe},
Addison-Wesley, Palo Alto, CA,;
Peebles, P. J. E. (1993) {\it Principles of Physical Cosmology},
Princeton University Press.
\item[5.]
Belinsky, V. A., Grishchuk, L. P., Khalatnikov, I. M.,
and Zeldovich, Ya. B. (1985) {\it Sov. Phys. JETP} {\bf 62}, 427.
\item[6.]
Grishchuk, L. P. (1987) {\it Sov. Phys. Uspekhi} {\bf 31}, 940.
\item[7.]
Harari, D. D. and Zaldarriaga, M. (1993) {\it Phys. Lett.} {\bf B319}, 96.
\item[8.]
Lidsey, J. E. (1993) Invited talk at the Second Alexander Friedmann
International Seminar on Gravitation and Cosmology;
Fermilab-Conf-336-A.
\item[9.]
Smoot, G. F. and Steinhardt, P. J. (1993) {\it Gen. Rel. Grav.} {\bf 25}, 1095.
\item[10.]
Bond, J. R., Crittenden, R., Davis, R. L., Efstathiou, G., and
Steinhardt, P. J. (1994) {\it Phys. Rev. Lett.} {\bf 72}, 13;
Turner, M. S. (1993) {\it Phys. Rev.} {\bf D 48}, 5539;
Atrio-Barandela, F. and Silk, J. (1994) {\it Phys. Rev.} {\bf D 49}, 1126;
Kolb, E. W. and Vadas, S. L. (1994) astro-ph. bulletin board 9403001,
Fermilab-Pub-94/046-A.
\item[11.]
Albrecht, A., Ferreira, P., Joyce, M., and Prokopec, T. (1994)
astro.ph bulletin board 9303001, Imperial College preprint TP/92-93/21.
\item[12.]
Grishchuk, L. P. (1993) {\it Phys. Rev. Lett.} {\bf 70}, 2371.
\item[13.]
Grishchuk, L. P. and Zeldovich, Ya. B. (1978) {\it Astron. Zh.} {\bf 55}, 209
[{\it Sov. Astron. {\bf 22}, 125 (1978)}].
\item[14.]
Lifshitz, E. M. (1946) {\it J. Phys. (JETP)} {\bf 16}, 587.
\vfill\eject
\item[15.]
Landau, L. D. and Lifshitz, E. M. (1975) {\it The Classical Theory of
Fields}, Pergamon Press.
\item[16.]
Bardeen, J. M. (1980) {\it Phys. Rev.} {\bf D 22}, 1882.
\item[17.]
Grishchuk, L. P. (1993) {\it Phys. Rev.} {\bf D 48}, 3513.
\item[18.]
Mukhanov, V. F., Feldman, H. A., and Brandenberger, R. H. (1992)
{\it Phys. Rep.} {\bf 215}, 6.
\item[19.]
Grishchuk, L. P. (1993) {\it Class. Quantum Grav.} {\bf 10}, 2449.
\item[20.]
Sakharov, A. D. (1965) {\it Zh. Eksp. Teor. Fiz.} {\bf 49}, 345
[{\it Sov. Phys. JETP {\bf 22}, 241 (1966)}];
Zeldovich, Ya. B. and Novikov, I. D. (1983) {\it The Structure and
Evolution of the Universe}, University of Chicago Press;
Peebles, P. J. E. (1981) {\it Astrophys. J.} {\bf 248}, 885.
\item[21.]
Grishchuk, L. P. (1992) in  H. Sato and T. Nakamura (eds),
{\it Proceed. VI-th M.Grossmann meeting on general relativity},
World Scientific, part~B, 1197.
\item[22.]
Misner, C. W., Thorne, K. S., and Wheeler, J. A. (1973)
{\it Gravitation}, W. H. Freeman \& Co., San Francisco.
\item[23.]
Grishchuk, L. P. and Sidorov, Yu. V. (1991)
in M. A. Markov, V. A. Berezin, and V. P. Frolov (eds),
{\it Procced. of V-th Moscow Seminar on Quantum Gravity},
World Scientific, p.~678.
\item[24.]
Landau, L. D. and Lifshitz, E. M. (1959) {\it Theory of Elasticity},
Pergamon Press.
\item[25.]
Brandenberger, R., Mukhanov, V., and Prokopec, T. (1993)
{\it Phys. Rev.} {\bf D 48}, 2443.
\item[26.]
Sachs, R. K., and Wolfe, A. M. (1967)
{\it Astrophys. J.} {\bf 1}, 73.
\item[27.]
Lukash, V. N. (1980) {\it Pis'ma Zh. Eksp. Teor. Fiz.} {\bf 31}, 631
[{\it Sov. Phys. JETP Lett. {\bf 6}, 596 (1980)}];
Chibisov, G. and Mukhanov, V. (1982) {\it Mon. Not. R. Astron. Soc.}
{\bf 200}, 535.
\item[28.]
Salopek, D. S. and Stewart, J. M. (1994) Hypersurface-invariant
approach to cosmological perturbations, DAMTP R94/25, ALBERTA THY/20-94,
astro-ph/9409055.
\item[29.]
Bardeen, J. M. Steinhardt, P. J. and Turner, M. S. (1983)
{\it Phys. Rev.} {\bf D 28}, 679.
\item[30.]
Starobinsky, A. (1982) {\it Phys. Lett.} {\bf B117}, 175;
Hawking, S. (1982) {\it Phys. Lett.} {\bf B115}, 295;
Guth, A. and Pi, S.-Y. (1982) {\it Phys. Rev. Lett.} {\bf49}, 1110;
Brandenberger, R. (1985) {\it Rev. Mod. Phys.} {\bf 57}, 1;
Brandenberger, B. (1987) {\it Int. J. Mod. Phys.} {\bf A 2}, 77.
\item[31.]
Linde, A., Linde, D., and Mezhlumian, A. (1994)
{\it Phys. Rev.} {\bf D 49}, 1783.
\item[32.]
Grishchuk, L. P. (1992) {\it Phys. Rev.} {\bf D 45}, 4717.
\item[33.]
Kashlinsky, A., Tkachev, I. I., and Frieman, J. (1994)
Fermilab-Pub-94/114-A.
\vfill\eject
\item[34.]
Grishchuk, L. P. (1994) Density perturbations of quantum-mechanical
origin and anisotropy of the microwave background,
WUGRAV-94-4, gr-qc/9405059 (to be published in Phys. Rev. D).
\end{itemize}
\end{document}